\documentclass{aa}        

\usepackage{xcolor}
\usepackage{graphicx}
\usepackage{txfonts}
\begin{document}

   \title{Two-Face(s): ionized and neutral gas winds in the local Universe.}


   \author{A. Concas
          \inst{1}\thanks{E-mail: alice.concas@tum.de (AC)}
          \and
          P. Popesso
          \inst{1}
          \and
          M. Brusa
          \inst{2,3}
             \and
        V. Mainieri
                 \inst{4}      
        \and 
        D. Thomas
        \inst{5}      
          }

   \institute{Excellence Cluster Universe,
                Boltzmannstr. 2  D-85748 Garching,  Germany 
         \and
         Dipartimento di Fisica e Astronomia,
              Universit\'a degli Studi di Bologna,
              Via Piero Gobetti 93/2, 40139 Bologna, Italy
             \and
             INAF$-$Osservatorio Astronomico di Bologna,
              Via Piero Gobetti, 93/3, 40139 Bologna, Italy
              \and
              European Southern Observatory, 
              Karl-Schwarzschild-str. 2, 85748 Garching, Germany
               \and
              Institute of Cosmology and Gravitation, 
              University of Portsmouth, 
              Dennis Sciama Building, Burnaby Road, Portsmouth, PO1 3FX, UK
             }

   \date{Received}

  \abstract
  {We present a comprehensive study of the Na I $\lambda$5890, 5895 (Na I D) resonant lines in the Sloan Digital Sky Survey (SDSS, DR7) spectroscopic sample to look for neutral gas outflows in the local galaxies. Individual galaxy spectra are stacked in bins of stellar mass ($M{\star}$) and star formation rate (SFR) {to investigate the dependence of galactic wind occurrence and velocity as a function of the galaxy position in the SFR-$M{\star}$ plane.}
While in most cases the interstellar medium (ISM) absorption and emission lines are at the galaxy systemic velocity, in massive galaxies ($M{\star} > 5 \times M_{\odot}$) at the high SFR tail (SFR$> 10-12 M_{\odot} yr^{-1}$) we find evidence of a significant blue-shifted Na I D absorption, {which we interpret as evidence of neutral outflowing gas. The occurrence of the blue-shifted absorption in the stacked spectra does not depend on the contribution of the nuclear activity, as it is observed at the same significance for purely star forming (SF) galaxies, active galactic nuclei (AGN) and composite systems at fixed SFR. 
We confirm, instead, for all classes of objects a clear dependence on the galaxy disk inclination: the blue-shift is the largest and the Na I D equivalent width the smallest for face-on galaxies while the absorption feature is at the systemic velocity for edge-on systems. This indicates that the neutral outflow is mostly perpendicular or biconical with respect to the galactic disk.} 
{We also compare the kinematics of the neutral gas with the ionized gas phase as traced by the [OIII]$\lambda$5007, H$\alpha$, [NII]$\lambda6548$ and [NII]$\lambda6584$ emission lines in the same galaxy spectra. 
Differently for the neutral gas phase, all the emission lines show evidence of perturbed kinematics only in galaxies with a significant level of nuclear activity and, they are independent from the galactic disk inclination. This would suggest that, while neutral winds originate from the galactic disk and are powered by SF feedback, ionized outflows are instead due to AGN feedback originating from the black hole accretion disk.}
{In both the neutral and ionized gas phases, the observed wind velocities (of the order of $100-200$ kms$^{-1}$) suggest that the outflowing gas remains bound to the galaxy with no definitive effect on the gas reservoir.}}


   \maketitle
%

\section{Introduction}

{Galaxies are not equally efficient in converting baryons into stars. It is now rather well established that the highest efficiency is reached by central galaxies of $10^{12-12.5}$ $M_{\odot}$ dark matter halos, like our own Milky Way, and it drops down quickly on both sides of this threshold towards lower and higher halo masses \citep[e.g.][]{Madau1996,Baldry2008,ConroyWechsler2009,Guo2010,Moster2010,Moster2013,Behroozi2010,Behroozi2013}. According to the current theoretical framework, outflows of different nature are the most viable way to expel the gas from a system and so regulate the star formation process through the gas availability \citep[e.g.][]{DiMatteo2005,DeLucia2006,Croton2006,Hopkins2006,Bower2006,Hopkins+14, Henriques+17}. In low mass halos ($< 10^{12-12.5}$ M$_{\odot}$) a combination of supernovae explosions and stellar winds in actively star-forming regions, is sufficient to swipe the gas away (see \citealp{Chevalier77}, \citealp{Murray+05} and \citealp{Hopkins+14} respectively for energy-driven outflows, momentum-driven outflows and for effect of multiple stellar feedback in cosmological simulations). In massive halos ($> 10^{12}$ M$_{\odot}$), instead,  the deeper potential well requires more energetic sources to expel the gas from the central system. Powerful feedback from active galactic nuclei (AGN) is believed to provide an effective mechanism in this respect \citep[]{DiMatteo2005,DeLucia2006,Croton2006,Hopkins2006,Bower2006,Hopkins+14, Henriques+17}, because the energy generated by the growth of the super-massive central black hole (BH) exceeds the binding energy of the gas by a large factor (see \citealp{Fabian2012} and \citealp{KingPounds2015}).}

{However, despite the ability of these models in predicting a large variety of evidence, a clear observational consensus regarding {\it{i)}} the efficiency of the different feedback as a function of stellar or dark matter halo mass and  {\it{ii)}}  their actual effect in regulating the star formation process is still lacking. Over the past decade, a lot of effort has been made to provide observational evidence of the existence and effect of the outflows. Star formation driven outflows are nearly ubiquitously observed in highly active star-forming galaxies at all cosmic epochs (see \citealp{Veilleux2005} and \citealp{Erb2015} for a comprehensive overview), usually associated with energetic starburst phenomena \citep[e.g.][]{Heckman+90,Pettini2000,Rupke2005a,Rupke2005b,Martin2005,Martin2006,HillZakamska2014}. Yet, their role in regulating the availability of gas is still largely debated \citep[e.g.][]{Martin+12,Rubin+14, Steidel+10}. Similarly, energetic outflows driven by AGN feedback are observed at low \citep[e.g.][]{Feruglio2010, Villar-Martin2011,RupkeVeilleux2011,Rupke+Veilleux13,Mullaney2013,Rodriguez-Zaurin+13,Concas+17} and high redshift \citep[e.g.][]{Maiolino2012,Brusa2015,Perna2015,Harrison+16,Zakamska+2016} with either positive (enhancement) or negative (suppression) effects on the galaxy star formation activity \citep{Cresci2015,Maiolino+2017,CresciMaiolino2018}.}

{It is now well established that the galactic winds are complex multi-phase phenomena, observed in neutral, ionized, and molecular gas phase in both 
SF galaxies \citep[e.g.][]{Heckman+90, Veilleux2005, Rupke2005a,Rupke2005b, Bordoloi+2014, Erb2015} and AGN dominated objects \citep[e.g.][]{Feruglio2010, Feruglio+2015, Villar-Martin2011,Harrison2014,Genzel+14,Brusa2015, Brusa2016, Concas+17}. However, there is still no general framework able to explain the relations between the different wind phases and their respective statistical incidence (see also \citealp{Cicone+2018}). This is likely because most observational evidence has been based so far on relatively small samples of peculiar objects as LIRGs and ULIRGs (Luminous and Ultra-Luminous Infrared Galaxies, e.g. \citealp{Heckman+90, Cazzoli+14,Cazzoli+16}) and massive AGN (\citealp[e.g.][]{GreeneHo2005a,Mullaney2013,ForsterSchreiber2018}) leading to strong selection biases. The low statistics often leads to inconclusive or contradictory results (see for example \citealp{Rupke+2017} and \citealp{Perna+2017a}). Only few attempts have been made to study the incidence and the role of outflows in the bulk of the galaxy population.
\cite{Concas+17} based on Sloan Digital Sky Survey (SDSS; \citealp{Abazajian2009}) stacked SDSS optical spectra, show that outflows of the ionized gas phase are taking place with a much higher incidence and velocity shift in galaxies with a strong AGN contribution and in particular at high stellar masses ($M{\star}$, see also \citealp{Cicone2016} for a similar study). \cite{Chen+10} use high signal to noise ratio (SNR) of SDSS stacked spectra to study the incidence of galactic outflows in the neutral gas phase through the Na I D absorption feature in the star forming galaxies population. They find a strong dependence of outflow incidence as a function of the disk inclination, which favors a biconical or perpendicular outflow geometry. However, no systematic study has been conducted so far to study the relative incidence and properties of different gas phase outflows in the bulk of the galaxy population.}

{In this paper, we conduct such systematic study by extending the analysis of the ionized gas phase of Concas et al. (2017) to the \textit{neutral} gas phase. For this purpose we use the spectroscopic sample of the SDSS at $z<0.3$ with a stacking analysis defined in Concas et al. (2017). As tracer of the neutral gas phase we use the Na I $\lambda$5890, 5895 (Na I D) resonant lines, which has been largely used to detect wind signature of blue-shifted absorbing material in front of the continuum source \citep[e.g.][]{Heckman2000,Rupke2005a,Rupke2005b,Rupke2005c,Martin2005,Rupke&Veilleux2015, Cazzoli+16, Rupke+2017}. Our goal is to characterize the kinematics of the neutral outflows in the bulk of galaxy population at low redshift to investigate {\it{i})} how they are related to the host galaxy properties, such as star formation rate (SFR), $M{\star}$ and galaxy geometry, {\it{ii})} identify the main triggering mechanism: star formation, AGN or a mixed contribution of the two effects and, {\it{iii})} explore if there is a difference between this cool neutral gas and the warmer ISM traced by the [OIII]$\lambda$5007, H$\alpha$, [NII]$\lambda6548$ and [NII]$\lambda6584$ emission lines in the same galaxies.}

This paper is organized as follows. Firstly, we describe our sample selection and physical properties in Section \ref{data}. The details of the method of creating stacked spectra and measuring the Na I D line profile are described in Section \ref{method}. We present and discuss the main results in Section \ref{results} and \ref{discussion}, respectively. Finally we summarize our findings in Section \ref{conclusions}. Throughout this paper, we adopt the following $\Lambda$CDM cosmology with $H_{0}=70$ km s$^{-1}$ Mpc$^{-1}$, $\Omega_{M}=0.3$ and $\Omega_{\Lambda}=0.7$

\section{The Data}\label{data}


   \begin{figure}
   \centering
   \includegraphics[angle=-90, width=\hsize ]{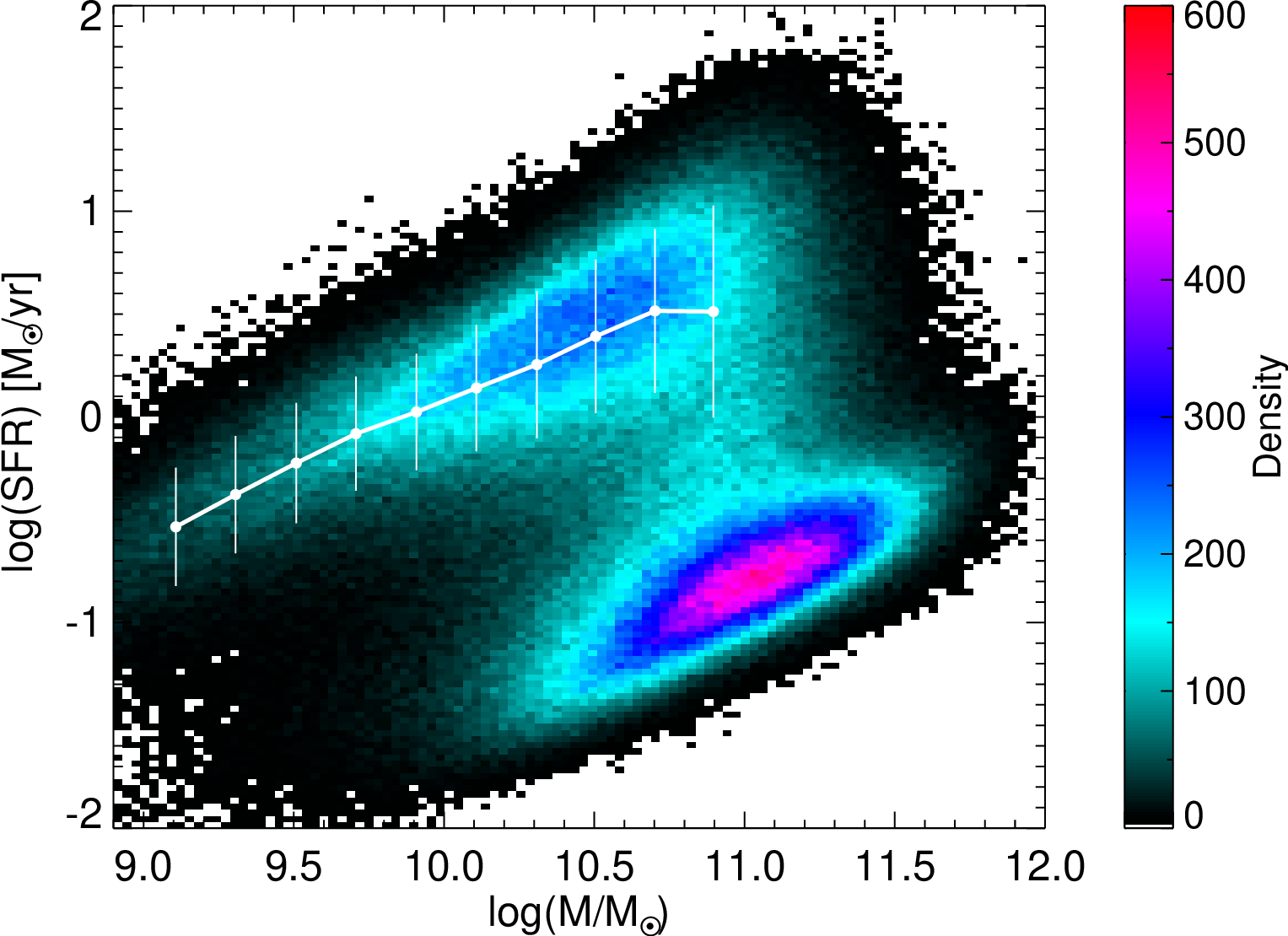}
   \caption{{The SFR vs. $M{\star}$ plane for our DR7 SDSS galaxies, color coded by the objects density as indicated in the color bar on the right. The white line shows the position of the so-called "Main-Sequence" (MS) of SF galaxies. The MS is computed as the mode and the dispersion of the SFR distribution in stellar mass bins following the example of \citep{RenziniPeng2015}.}}
              \label{FigPlane}%
   \end{figure}

The dataset used in the present study is based on the SDSS (\citealp{York2000}) spectroscopic catalog, data release 7 (DR7, \citealp{Abazajian2009}). In particular, we analyze objects drawn from the Main Galaxy Sample (MGS, \citealp{Strauss2002}) which have Petrosian magnitude $r<17.77$ and redshift distribution spanning $0.005 < z < 0.30$. The spectra are obtained with $3''$ diameter fibers, that cover a wavelength range from $3800 < \AA < 9200$, with an instrumental resolution of $R \equiv\lambda / \delta \lambda \sim 1850-2200$ and a mean dispersion of $69$ km s$^{-1}$ pixel$^{-1}$. See http://www.sdss.org/dr7/ for more exhaustive details concerning the DR7 spectra. 

{Throughout the paper, we use the $M{\star}$ and SFR estimates of the the MPA-JHU catalogue, which provides derived galaxy properties for more than $800000$ galaxies\footnote{http://www.mpa-garching.mpg.de/SDSS/DR7/}.} Briefly, the stellar masses are estimated by fitting the broad-band optical photometry (\citealp{Kauffmann2003a}) and the SFR measurements are based on the \cite{Brinchmann2004} procedure.  The H$\alpha$ emission line luminosity is used to determine the SFRs for the SF galaxies and the D4000-SFR relation \citep[e.g.][]{Kauffmann2003a} for all galaxies that have emission lines contaminated by AGN activity or non detected. This leads to a SFR estimate for all objects. All SFR measures are corrected for the fiber aperture as described by \cite{Salim2007}. 
{We restrict the sample to stellar mass $\log(M/M_{\odot}) \geq 9.0$ to limit the incompleteness in the low mass regime. The galaxy sample is shown in the SFR-stellar mass plane in Fig.\ref{FigPlane}, and it contains $\sim 600,000$ (see Concas et al. (2017, hereafter C17, for further details on the sample description).}

The MPA-JHU catalog also includes the flux values of several emission lines for each galaxy. In order to distinguish normal SF from AGN dominated galaxies we classify galaxies in the \citeauthor*{BPT}(1981, BPT) diagnostic diagram on the basis of the line ratios: [OIII]$\lambda$5007/H$\beta$ and [NII]$\lambda$6584/H$\alpha$. 
{The use of the [NII]$\lambda$6584/H$\alpha$, in particular, allow as to carefully discriminate between the ionization does to the presence of photons originate from SF and AGN activity.
As reported by previous studies such emission line ratio is more sensitive to the presence of
low-level AGN than other diagnostics such as [SII]/H$\alpha$ and [OI]/H$\alpha$ (see, e.g. \citealp{Kewley2006,Schawinski+2010}).}
Following the \cite{Stasinska2006} classification, we define subsamples on the basis of the prevalence of different photoionization processes: SF, SF$-$AGN, AGN$-$SF, LINERs and TYPE 2, respectively 20.6\%, 7.4\%, 11.2\%, 5.6\% and 1.7\% of the total sample. In this paper the LINERs and TYPE 2 galaxies are classified together as AGN galaxies.
All galaxies with no or very weak emission lines ($SNR<4$), 53\% of the total sample, are not classifiable using the BPT diagram and we call these objects "un-Class". {As already mentioned in the Introduction, the main goal of this study is 
to investigate and quantify the evidence of neutral gas outflows in the bulk of
galaxies population in the local universe, therefore the inclusion of such non-emitter galaxies is mandatory in the following analysis. However, we checked that our main results are not affected by the inclusion of such "un-Class" galaxies.}

\begin{figure}
\centering  
\includegraphics[angle=90, width=\hsize]{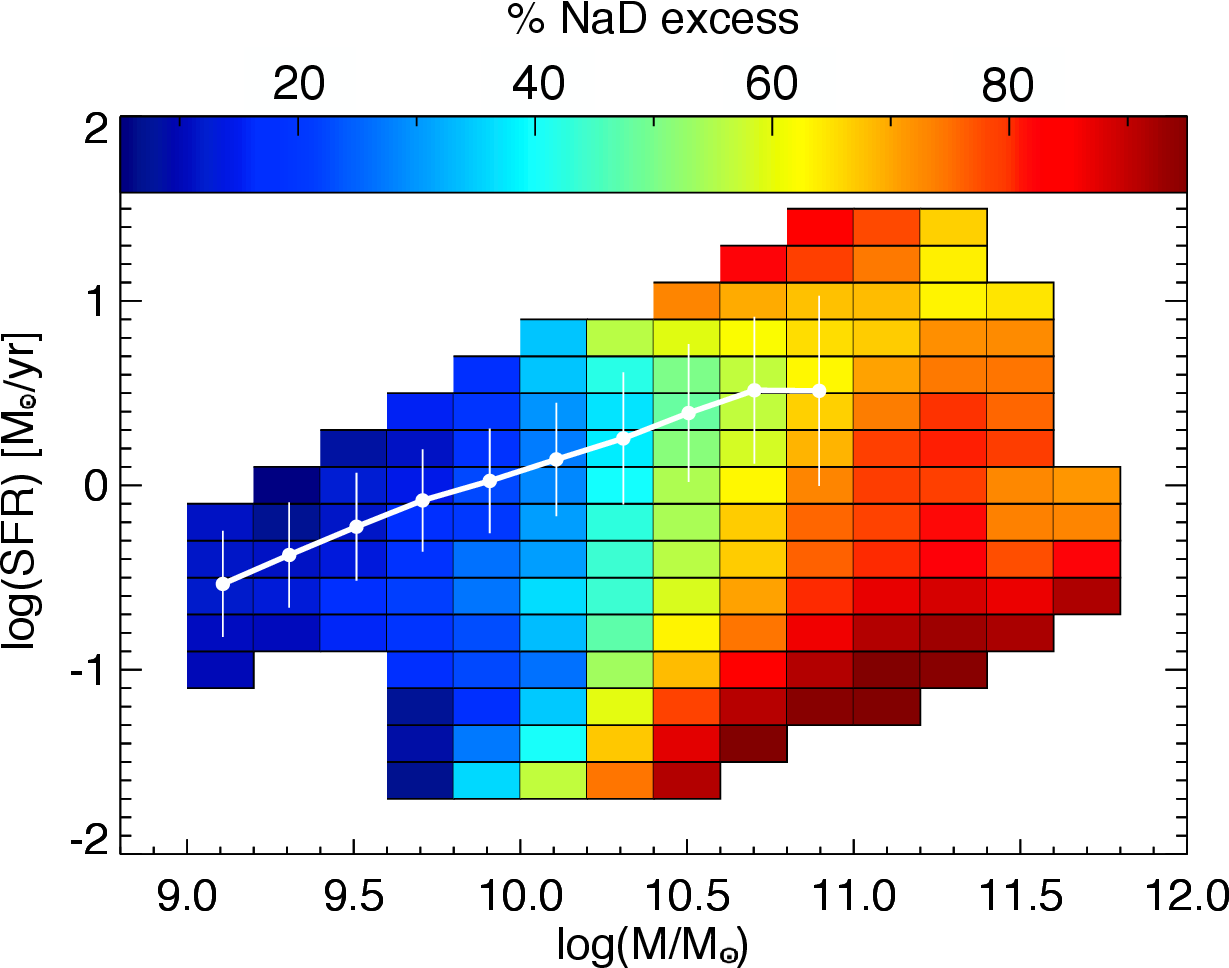}
\caption{{The SFR vs. $M{\star}$ plane for our DR7 SDSS galaxies, color coded by the percentage of objects showing a Na I D excess (NaD$_{obs} >$ NaD$_{mod}$) in each bin, from the Na I D measurements of the MPA-JHU catalog (only object with SNR >2 in the single observed Na I D Lick measurements). The Na I D excess objects dominate the high M$_{\star}$ region, while the number of the Na I D deficiency (NaD$_{obs} <$ NaD$_{mod}$) objects increases at low stellar masses. The white line shows the mode and dispersion of the MS (see Fig. 1).}}

\label{fig:FigNEO_NDO}
\end{figure}   

\begin{figure*}
\centering
\includegraphics[angle=0, width=0.49\textwidth]{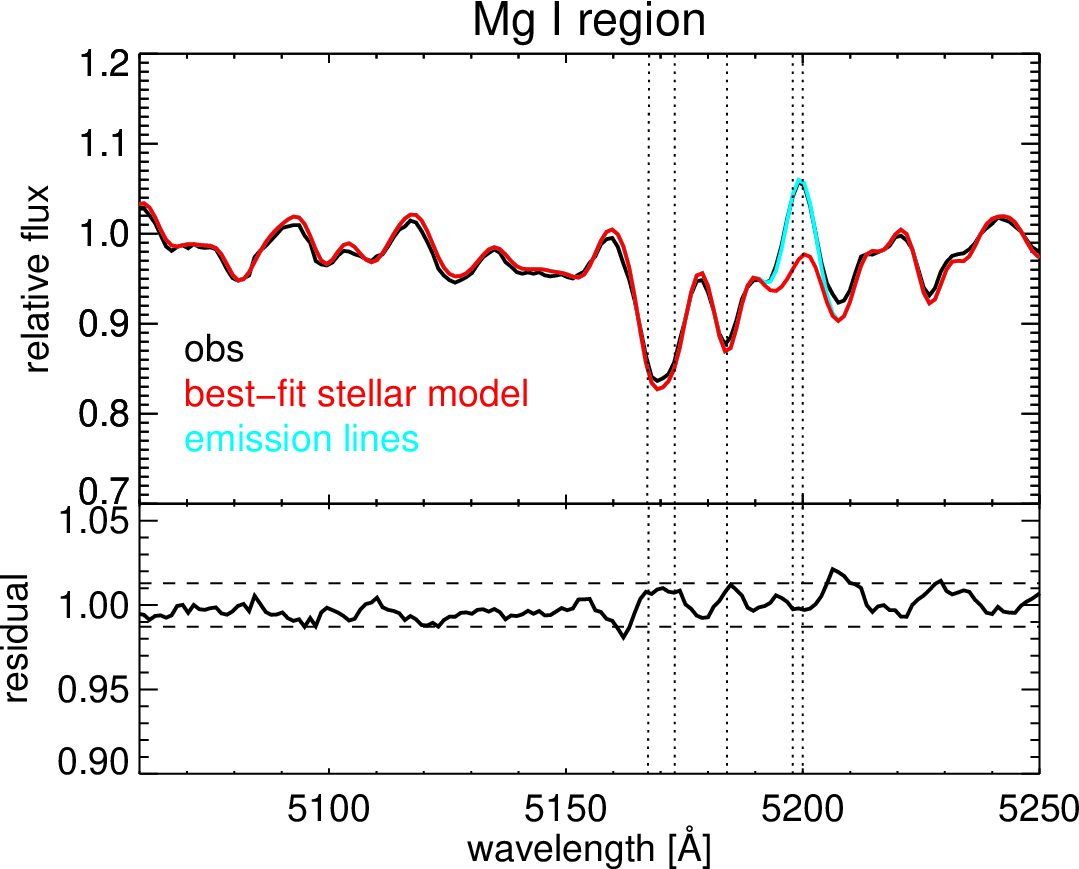}
\includegraphics[angle=0 , width=0.49\textwidth]{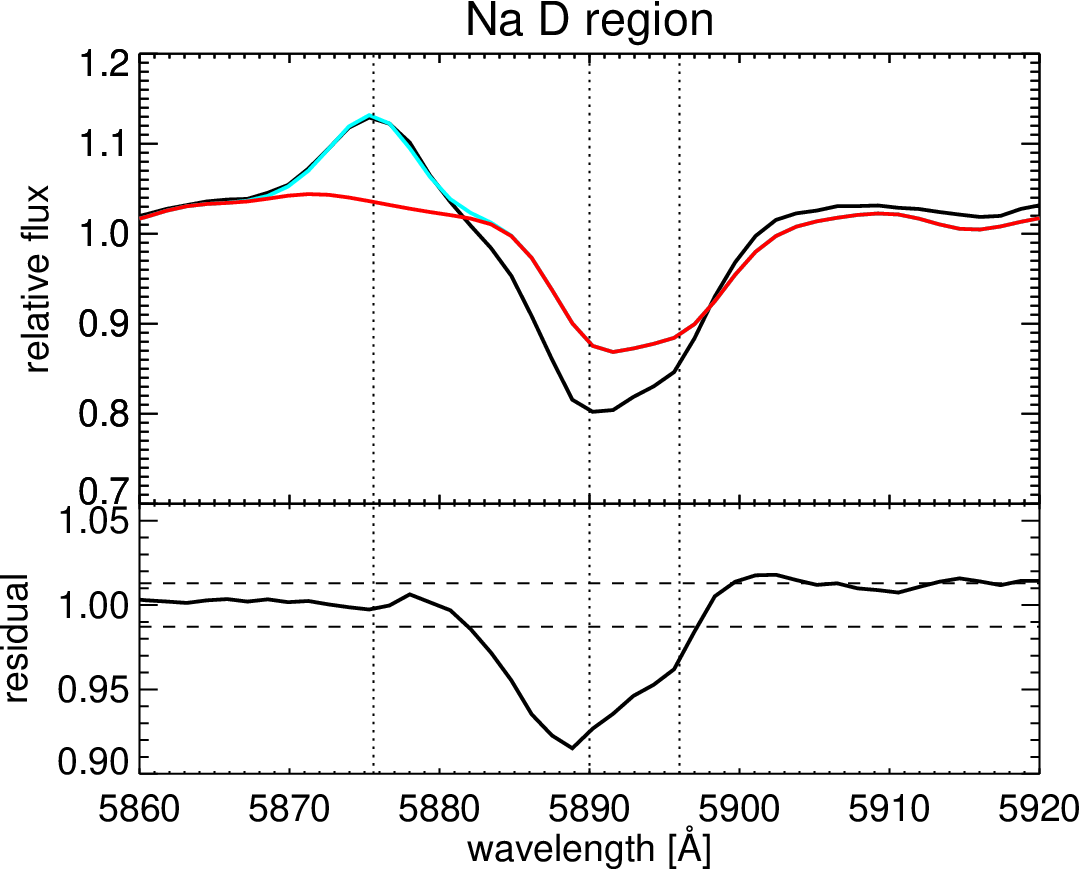}
\caption{Example of our continuum fit and subtraction performed for the stacked spectrum whit $\log$(SFR/M$_{\odot} yr^{-1}$)=1.4 and $\log$(M$_{\star}/M_{\odot}$)=10.9 in the Mg I and Na I D regions, left and right panel respectively. The observed stacked spectrum is shown with a solid black line, the best-fit stellar continuum model in red and stellar emission lines in cyan. The dotted lines indicate the stellar Mg I triplet at $\lambda\lambda$5167, 5173, 5184 \AA\, the [NI]$\lambda$5198, 5200 \AA\ emission lines (left panel), the He I $\lambda$5876 emission and the Na I doublet $\lambda\lambda$5890, 5896 (right panel).
   {The bottom panels shows the residual spectra (solid black line) and the residual variations (dashed lines).} The Mg I region is well reproduced by the stellar fit with residuals lower than the mean scatter. The Na I D fit is instead affected by significant residuals, revealing a blue-shifted extra absorption component.}
\label{fig:fit_MgI_NaDregions}
\end{figure*}   
\subsection{ISM Na I D in galaxy spectra}\label{singoli}
{In this work we focus on the Na I doublet at $\lambda\lambda$5890,5896 \AA\ \footnote{The transition which gives rise to the doublet is from the ground-state configuration 3s to the energy levels of the 3p configuration composed by two states with total angular momentum j=3/2 and j=1/2 (spin-orbit effect). The continuum photons emitted by the stellar component are absorbed by the Na I atoms (at temperatures T < 1000K) and are re-emitted when the electrons in the 3p configuration decay spontaneously to the ground-state.} resonant lines in the rest-frame optical galaxy spectra. It is now generally recognized that more than one process causes the formation of this transition in a galaxy spectrum, including 1) the absorption and re-emission of stellar radiation by the ISM as well as 2) the absorption of radiation produced in the stellar interiors by the most external gas layers (stellar contribution). It is clear that to identify any signature of outflow in the ISM component, this must be isolated through the removal of the stellar component. This can be done by fitting the stellar continuum so to derive the stellar contamination of the Na I D feature. This analysis requires very high SNR spectra {\it{i})} to properly fit the stellar continuum and {\it{ii})} to obtain a high SNR residual ISM component in order to derive the gas kinematics. Only the stacking analysis of the SDSS spectra allows to reach such high SNR. However, this poses a problem. The ISM component can be in emission or absorption, with an incidence strongly dependent on the dust obscuration (\citealp{Chen+10}). The emission and absorption in the stacking of a spectral feature can easily cancel out each other and so lead to very biased results. Thus, before proceeding with the stacking procedure described in the next section, we first verify the incidence of emission and absorption of the Na I D interstellar component in the SFR-$M{\star}$ plane.}

{This test is done by using the information contained in the MPA-JHU catalog\footnote{http://wwwmpa.mpa-garching.mpg.de/SDSS/DR7/SDSS$\_$indx.html}, which provides a measure of the Na I D line strength parametrized by the Na I D Lick index (as defined in \citealp{Worthey+1994}).  The catalog includes a measure of the Na I D Lick index due to the stellar component (NaD$_{stellar\_mod}$), derived by fitting the continuum as described in \cite{Tremonti+04}, together with a measure of the observed Na I D Lick index NaD$_{obs}$, which is given by the combined ISM and stellar contribution. A $NaD_{obs} > NaD_{stellar\_mod}$ (hereafter Na I D excess) implies an extra ISM absorption component, while a $NaD_{obs} < NaD _{stellar\_mod}$ (hereafter Na I D deficiency) points to a ISM emission component. The distribution of Na I D excess and deficiency is not perfectly Gaussian, as outlined by Chen et al. (2010), but it exhibits a long tail on the positive side (excess). We investigate the incidence of ISM emission and absorption in the SFR-$M{\star}$ plane only for the SDSS subsample, where NaD$_{obs}$ is observed at SNR$>2$. As shown in fig. Fig.\ref{fig:FigNEO_NDO}, the Na I D excess and deficiency are not homogeneously distributed in the plane, but they exhibit a bimodal distribution as a function of the galaxy stellar mass. The deficiency is dominating (80\%) at low stellar massive and the excess is dominating at the high mass end. Only in the transition region at $10^{10-10.5}$ $M_{\odot}$, emission and absorption are mixed together in similar percentages. This is consistent with recent high redshift galaxy studies of the ISM FeII and MgII ISM component in the rest-frame near-UV, tracing the same gas phase (see \citealp{Finley+2017,Feltre+2018}). The observed bimodal distribution of the Na I D deficiency and excess allows us to perform the stacking analysis in the SFR-M$\star$ plane with the caveat that in the central region, at $M_{\star}=10^{10-10.5} $ M$_{\odot}$ the result might be biased.}

\section{Method}\label{method}
In this section, we describe how we fit and remove the stellar absorption contribution and ionized emission lines and how we measure the properties of the ISM Na I D resonant line in our stacked spectra. 

\subsection{Stacked spectra and stellar continuum fit} 
{In order to improve the SNR of our single spectra and, at the same time, to connect the Na I D feature to the galaxy properties (SFR, $M{\star}$ and main ionization source), we stack together the optical spectra of the C17 sample in bins of SFR and $M{\star}$ ($\Delta \log(M/M_{\odot}) =0.2$ and $\Delta \log(SFR) =0.2$ dex). 
Following the approach of C17, we obtain $148$ galaxy stacked spectra in the C17 sample all over the SFR-M$\star$ plane (see the fine grid in Fig. 1 in C17).}


We use a combination of two publicly available codes: pPXF \cite{Cappelalri_Emsellem2004} to fit the stellar continuum , and GANDALF  \cite{Sarzi2006} to fit the nebular emission lines. As in C17 we use the BC03 stellar library (\citealp{BC03}) to produce a model of the stellar continuum that matches the observed line-free continuum in the wavelength range $3800 < \lambda <6900$ \AA. Our templates include simple stellar population with ages $0.01 \leq t \leq 14$ Gyr, four different metallicities, $Z/Z_{\odot}=0.2, 0.4, 1, 2.5$ and assume a \cite{Chabrier2003} initial mass function (IMF). {{Any interstellar medium emission and absorption lines, including Na I D, is ignored in the continuum fit. The emission lines are separately fitted with GANDALF. The two contributions, continuum$+$emission lines, are then subtracted from the observed spectrum to isolate the ISM component. Fig \ref{fig:fit_MgI_NaDregions} shows an example of the stellar continuum (red curve) and emission line (cyan curve) fit versus the observed spectrum (black curve) in two regions,  the Mg I (left panel ) and Na I D (right panel). In both plots, the bottom panels show the residual after the subtraction of continuum$+$emission lines. While the Mg I triplet at $\lambda$5167,5173,5185, due only to stellar atmosphere, is very well fitted (it is known that EW(NaD)$_{star} =0.75 \times$ EW(MgIb)$_{star}$ \citealp{Heckman2000}, see also \citealp{Rupke2002,Rupke2005a,Villar-Martin2014}), the Na I D region exhibits the expected extra absorption of the ISM component.

In the right panel, in particular, we highlight the importance of removing the He I $\lambda$5876 emission line in order to properly isolate the Na I D ISM contribution. Such emission is doe to the presence of energetic photons able to photoionize the He atoms in the HII regions in SF galaxies (see also \citealp{Chen+10}).
For such galaxies spectra the He I line is fit with a single Gaussian component and then removed from our spectra. We check that a multi-gaussian fit is not required for this emission line. This "residual" spectrum (bottom right panel in Fig. \ref{fig:fit_MgI_NaDregions}) is finally used for any analysis of Na D line features.}}

The error of the "residual" spectrum is obtained through bootstrap re-sampling methods shown in C17.

\subsection{Measuring the ISM Na I D absorption line profiles}\label{ISMmeasure}
In our stacked spectra, the velocity structure of the Na I D line is generally complicated; this resonant line shows absorption, emission, or a combination of two contribution across the galaxies physical parameters (as also found in the SF galaxy spectra showed in \citealp{Chen+10} and in the nearby quasars in \citealp{Rupke&Veilleux2015,Rupke+2017}).
To correctly quantify the centroid velocity and velocity dispersion of the Na I D resonant line, a careful treatment of the absorption and emission lines is mandatory. 
Each stacked spectrum is fit with three distinct models using the IDL MPFIT fitting code.\\
\\

{Model 1.} The first model, $one-free-doublet$ model, consists of a pair of Gaussians with a single kinematic component (as in previous works, e.g. \citealp{Davis+12,Cazzoli+14,Cazzoli+16}). 
%
%
%
The four free parameters of the model are the line centroid of the strongest blue absorption in the doublet ($\lambda_{1}$), the line width ($\sigma_{1}$), the amplitude of the blue component (A$_{\lambda5890}$), and the ratio between the amplitudes of the blue and red components (A$_{\lambda5890}$/A$_{\lambda5896} = \alpha$).
The central blue wavelength of the Na I D doublet (Na I D $\lambda5890$), $\lambda_{1}$ is tied to define a single Doppler shift to respect the galaxy's systemic velocity (defined as the systemic redshift, $\lambda = 5890$ \AA), $\Delta V_{1}$.The ratio between the two amplitudes, is limited to vary between an optically thin $\alpha$=2 and optically thick absorbing gas, $\alpha$=1 \cite{Spitzer1978}.\\

This approach allows us to characterise the kinematics of the global observed doublet without introducing other model dependent parameters. However, some stacked spectra show complex profiles, that suggest the presence of an additional Gaussian pair, in emission or in absorption (e.g. \citealp{Chen+10,Cazzoli+16}).
This is motivated by two main facts.
First, the scattering processes that arise in the flowing material can produce a P Cygni-like profiles of Na I D (see the analysis of the absorption and emission-line profiles for the Mg II $\lambda2796,2803$ doublet and Fe II multiplet at $\lambda 2600$ \AA\ in cold gas wind models shown in \citealp{Prochaska+11}), characterized by blue-shifted absorption plus resonance line emission at roughly the systemic velocity.  

To model both the P Cygni profile and the interstellar absorption or emission at the systemic velocity, we introduce a second pair of Gaussians at the $one-free-doublet$ model.\\
In particular we define two new models of double doublets: the $two-free-fix-doublet$ and the $two-free-doublet$, respectively, model 2 and 3.\\
\\
{Model 2.} The $two-free-fix-doublet$ consist of two gaussian doublets: one fixed to the systemic velocity and one free to varies in the wavelength space. The model is described by: the line centroid of the two doublets ($\lambda_{1}$ and $\lambda_{2}$ ), the lines widths ($\sigma_{1}$ and $\sigma_{2}$), the amplitudes of the blue lines (A$_{\lambda5890}{1}$ and A$_{\lambda5890}{2}$) and the ratio $\alpha_{1}$ and $\alpha_{2}$.
We note that often the $two-free-doublet$ leads to unphysical solutions, since the absorption or emission troughs are not well resolved. In order to not incur in this spurious results, the width of the second pair of Gaussians is constrained to the velocity dispersion of the nebular gas previously fitted to the H$\alpha$ emission line (this is similar to those assumed in previous studies e.g. \citealp{Cazzoli+16} and \citealp{Martin2012} by using the [OII] doublet). The model described the observed line with five free parameters.
As in the first model the flux ratio ($\alpha_{1}$ and $\alpha_{2}$) is set to vary between 1 and 2.
The velocity of the outflowing gas is now expressed by the shift of the central blue wavelength ($\lambda_{1}$), $\Delta V_{2}$. \\
\\
{Model 3.} The $two-free-doublet$ is equal to the model 2 but with both the doublets free to vary in the wavelength range (seven free parameters). In this way, also the non systemic red component (in emission or absorption) is well fitted. The blue-shift of the outflowing component is referred to as $\Delta V_{3}$. \\
\\
{To avoid over-fitting and to allow the correct number of Gaussian doublets used in each Na I D line fit, we employ the the Bayesian Information Criterion (BIC, \citealp{Liddle2007}). BIC is a likelihood criterion penalized by the model complexity widely used in the problem of model identification. This method is particularly useful when a set of different candidates models with different number of parameters is used to describe a given data set. 
The BIC criterion is defined as: $BIC= \chi^{2} + p * \ln(n)$ , where $\chi^{2}$ is the chi squared of the fit, $p$ is the number of free parameters and $n$ is the number of flux points used in the fit. We measure the BIC value by using the three fitting methods described above, BIC1, BIC2 and BIC3, respectively. The model with the smallest value of the BIC is chosen as the preferred model for the data. We visually inspect each fit to evaluate the accuracy of this procedure. According to this statistical method we use $\Delta V_{1}$, $\Delta V_{2}$ and $\Delta V_{3}$ and $\sigma_{1}$,$\sigma_{2}$ and $\sigma_{3}$ of the fit with the lowest BIC value to quantify the blue-shift and the width of the outflowing component.} \\
\\
In order to estimate measurements uncertainties, the observed spectrum is perturbed randomly accordingly to the observed SNR. We simulate, in this way, 1000 different realizations of the same observed spectrum.  We fit the obtained noisy spectra with our 3 methods. Each quantity ($\Delta V_{1}$, $\Delta V_{2}$ , $\Delta V_{3}$, $\sigma_{1}$,$\sigma_{2}$ and $\sigma_{3}$) is measured in all perturbed template spectra. The 68$\%$ confidence interval of each quantity distribution around the mean value is adopted as a measure of the uncertainty.

As in \cite{Rupke2005b}, we define a blue-shifted detection only if the difference between the velocity peak and the systemic velocity is  $> 70$ kms$^{-1}$ (instrumental resolution of our SDSS data).  To better parameterize the blue wings of the neutral Na I D velocities, we also provide a measure of the maximum velocity, $V_{max}= \Delta V - 2\sigma $ as defined for the absorption Na I D component in \cite{Rupke&Veilleux2015}.


\section{Results}\label{results}

\subsection{Na I D lines in the global sample}\label{globalSample}
\begin{figure}
   \centering
   \includegraphics[angle=90, width=\hsize ]{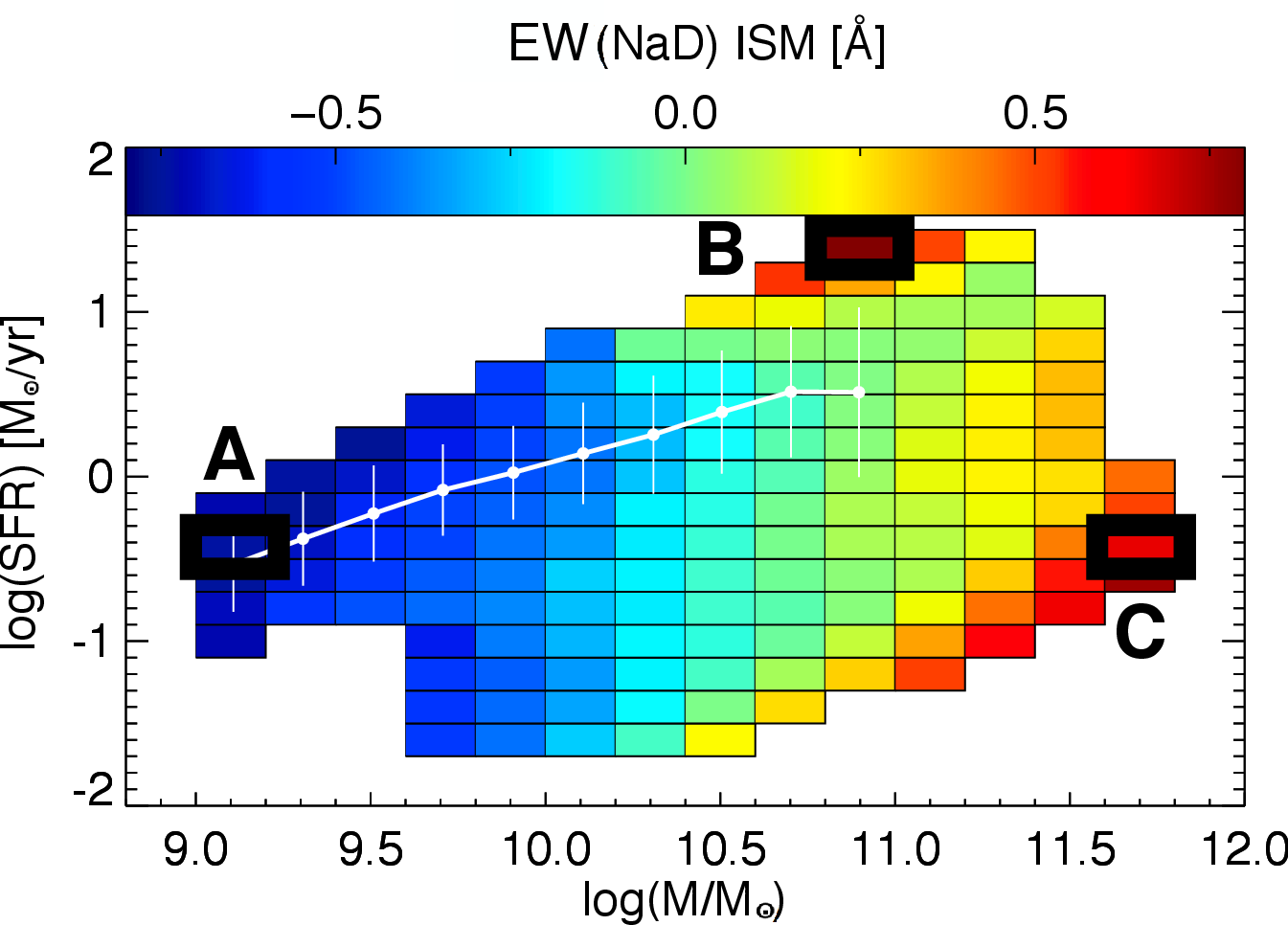}
      \caption{The SFR vs. $M{\star}$ plane for our DR7 SDSS galaxies, color coded by the intensity of the EW of the ISM NA I D line
{after the stellar continuum subtraction. Negative and positive EW values correspond to a residual emission and absorption feature, respectively. 
 The white line shows the mode and dispersion of the MS (see Fig. 1).
For the highlighted Bins, A, B and C,  the Na I D line fit is shown in Fig. 5.}}
              \label{Fig_eqw_SNtot}
   \end{figure}

\begin{figure}
   \centering
   \includegraphics[ width=\hsize ]{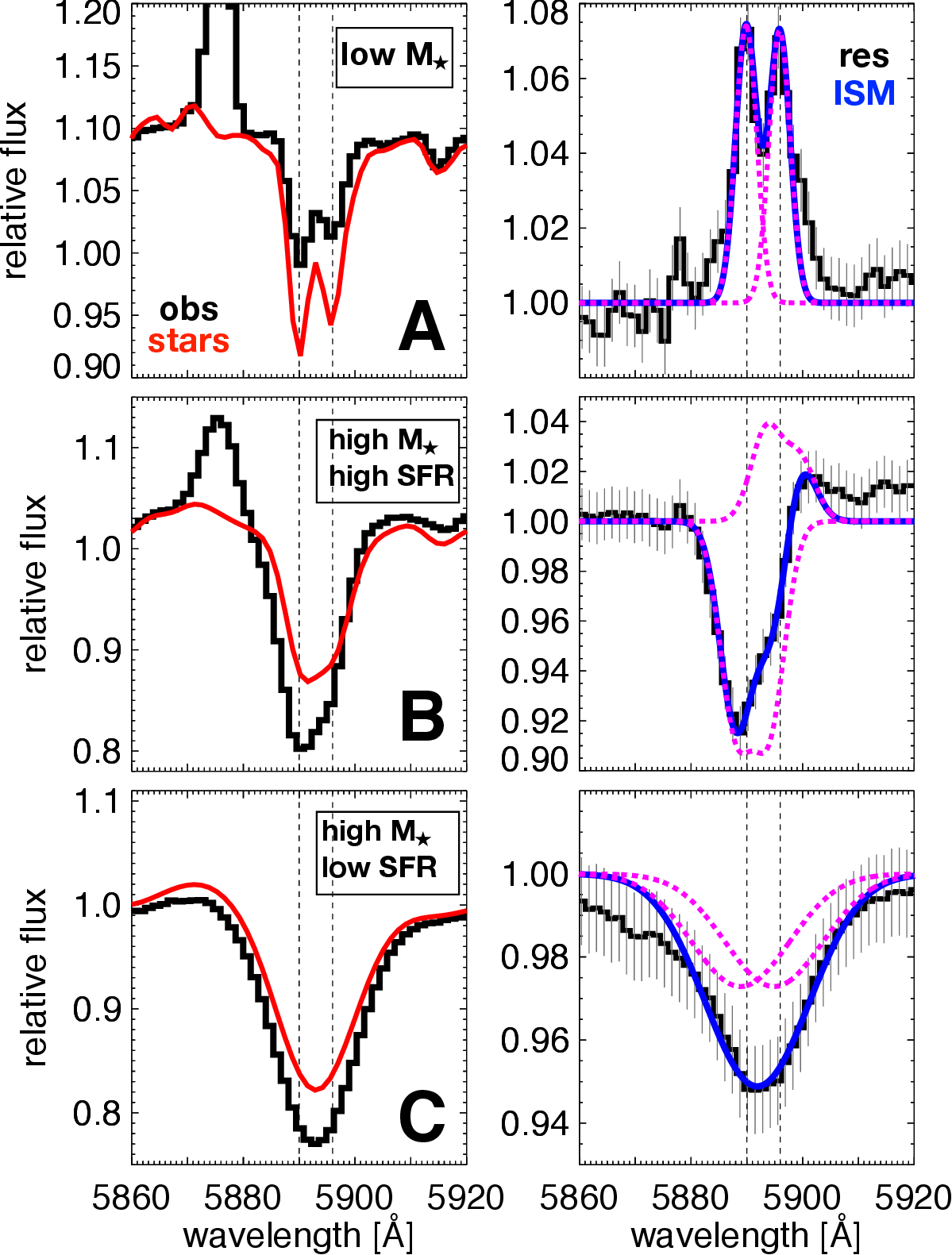}
   \caption{{{Examples of Na I D fits of the stacked spectra from three different regions of the SFR-M$_{\star}$ plane, labeled in Fig, 4: low stellar mass ( $\log$(M)$_{\star} < 10.5 M_{\sun}$, top panels, bin A in Fig. 4), high stellar mass and high SFR ($\log$(M)$_{\star}> 10.5 M_{\sun}$ $\log$(SFR)$>1.1$ M$\odot$/yr, central panels, bin B in Fig. 4) and low SFR and high stellar mass ($\log$(M)$_{\star}> 10.5 M_{\sun}$, $\log$(SFR) $< 0.5$ M$\odot$/yr, bottom panels, Bin C in Fig. 4). The expected wavelengths of Na I D are shown as vertical dashed lines. {\it{Left panels}}:  }}
  { The black histogram illustrates the observed stacked spectrum, the red line the best-fit stellar continuum.  {\it{Right panels}}:  The black histogram shows the residual flux and errors after the stellar continuum has been removed. The best-fit of the doublet is shown in blue. The dotted magenta curves are the single Gaussian components. Only in bin B, high stellar mass and high SFR bin (central panel), we detect a doppler blue-shift of the Na I D absorption feature with  $\Delta V = -102 \pm 41$ km s$^{-1}$. In the other two bins the Na I D line shows a zero-velocity emission (bin A, top panel) and absorption (bin C, bottom panel), below the instrumental resolution $70$ km s$^{-1}$. }}
              \label{Fig_3Region}
   \end{figure}

After subtraction of the stellar continuum, we measure the residual equivalent width, EW(NaD), in all the 148 stacked spectra of the C17 Sample. Fig. \ref{Fig_eqw_SNtot} shows the EW(NaD) obtained for each stacked spectrum in the SFR$-M_{\star}$ plane. {{As found for individual galaxy spectra in Fig.\ref{fig:FigNEO_NDO} , also for the stacked spectra the mean EW(NaD) increases progressively from negative values (emission) to positive values (absorption) as a function of the $M{\star}$ and with no dependence on the SFR up to $\sim 4-5\times 10^{10}$ $M_{\odot}$. At the high mass end ($> 5\times 10^{10}$ $M_{\odot}$), we observe also a different EW(NaD) as a function of the SFR. Indeed, EW(NaD) reaches its maximum in the starburst region at high SFR and in the quiescent region at very low SFR, respectively with a mean value of EW(NaD)$\sim 0-0.25$ $\AA$ on the Main Sequence and in the transition region. By looking at the individual spectra it is possible to identify key differences in the Na I D line profile. 

We applied the fitting procedure described in Section \ref{ISMmeasure} to all the 148 stacked spectra in order to derive the neutral gas kinematics expressed through the $\Delta V$ and $\sigma$ parameters.
To facilitate the view, we isolate three examples of representative regions of the diagram: one at very low stellar masses (region A), one at high stellar mass and SFR (region B) and one at high stellar mass and low SFR (region C). The Na I D region of the three spectra is shown in Fig. \ref{Fig_3Region}. In particular, the left panels show the observed spectrum (black curve) and the best fit stellar contribution (red curve), while the right panels show the ISM residual obtained after removal of continuum$+$emission lines. As previously reported by \cite{Jeong+13}, we find that low mass galaxies, at stellar masses below $\sim 4-5\times 10^{10}$ $M_{\odot}$, at any SFR are characterized by a residual Na I D emission at the galaxy systemic velocity (top panels of Fig. \ref{Fig_3Region}), likely arising from HII regions (e.g. \citealp{Bica1986})}. At higher stellar masses, $M{\star}$ > $5\times 10^{10}$ $M_{\odot}$ and {\it{high SFR}}, the EW(NaD) increases to positive values and shows an indication of a slightly blueshifted absorption (central panels of Fig. \ref{Fig_3Region}). At high M$\star$ and low SFR, in the quiescent region (region C), we do observed a residual Na I D absorption centered at the galaxy systemic velocity (bottom panels of Fig. \ref{Fig_3Region}).} The existence of a Na I D residual absorption in the early-type galaxies is known (e.g. \citealp{Carter+1986, Alloin1989, Thomas+2003, Worthey1998,Worthey2011}) and has been largely debated. This could be due to {\it{i})} an extra Na abundance in the ISM, or {\it{ii})} in the stellar atmospheres (e.g. \citealp{O'Connell1976,Peterson1976,Parikh+2018}) or {\it{iii})} to a bottom-heavy stellar initial mass function (IMF) \citealp{vanDokkum+10, vanDokkum+12,Spiniello+12}). We confirm here that in any of the stacked spectra of the quiescence region the extra Na I D absorption is observed at the systemic velocity and it does not show any signature of ongoing outflow.

   \begin{figure}
   \centering
   \includegraphics[angle=0, width=\hsize ]{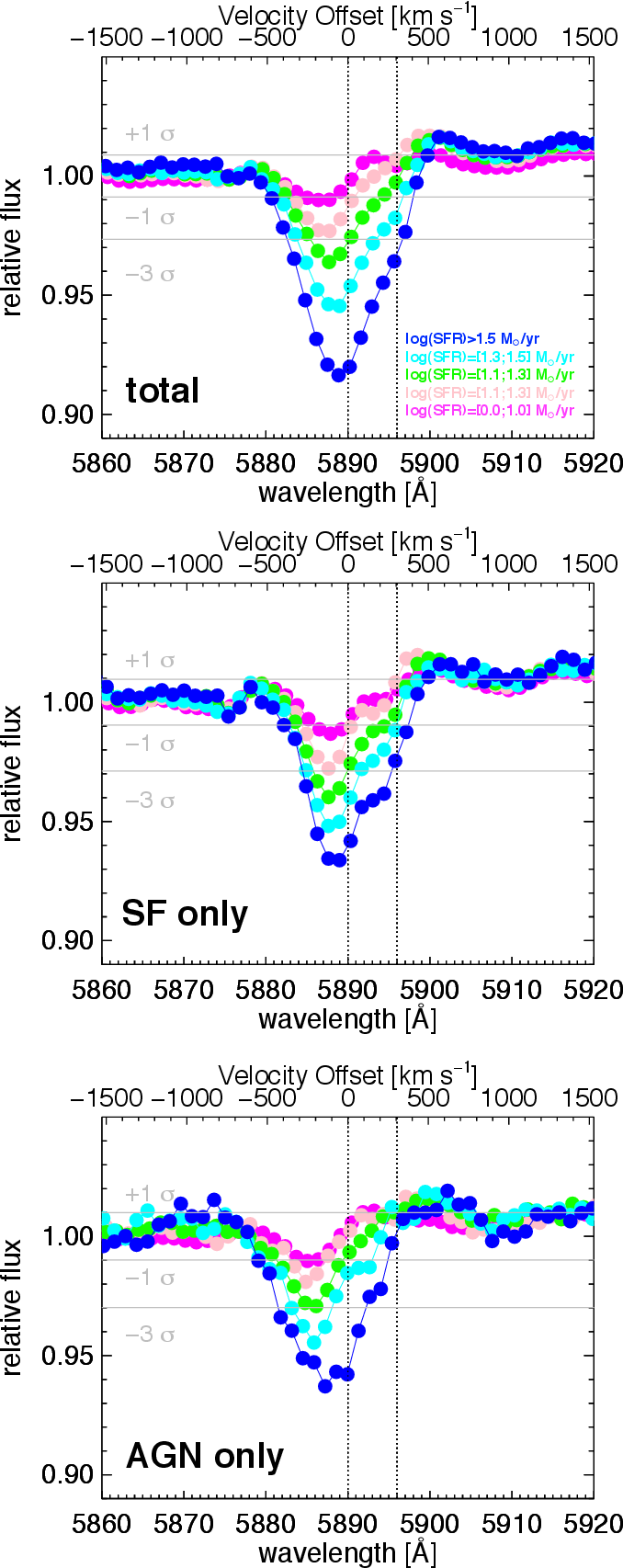}
   \caption{Variation of the ISM Na I D resonant line profile as a function of the SFR for the galaxies with $\log$(M)$_{\star} > 10.7$ M$_{\odot}$ and $\log$(SFR)$ > 0.0$ M$_{\odot}$/yr in the total sample (top panel), {in ``pure'' SF galaxies (middle panel)} and in AGN dominated objects (bottom panel). The intensity of the line increases with increasing SFR, from the magenta to the blue curves. Only galaxies with SFR$\geq 12$ M$_{\odot}$/yr show a blue-shifted Na I D line, detected with a significance $>3 \sigma$. 
{The ``pure'' SF and AGN subsamples show the same trend observed in the total sample.}}
              \label{NaD_SFR}
   \end{figure}

   \begin{figure*}
   \centering
   \includegraphics[angle=-90, width=\hsize ]{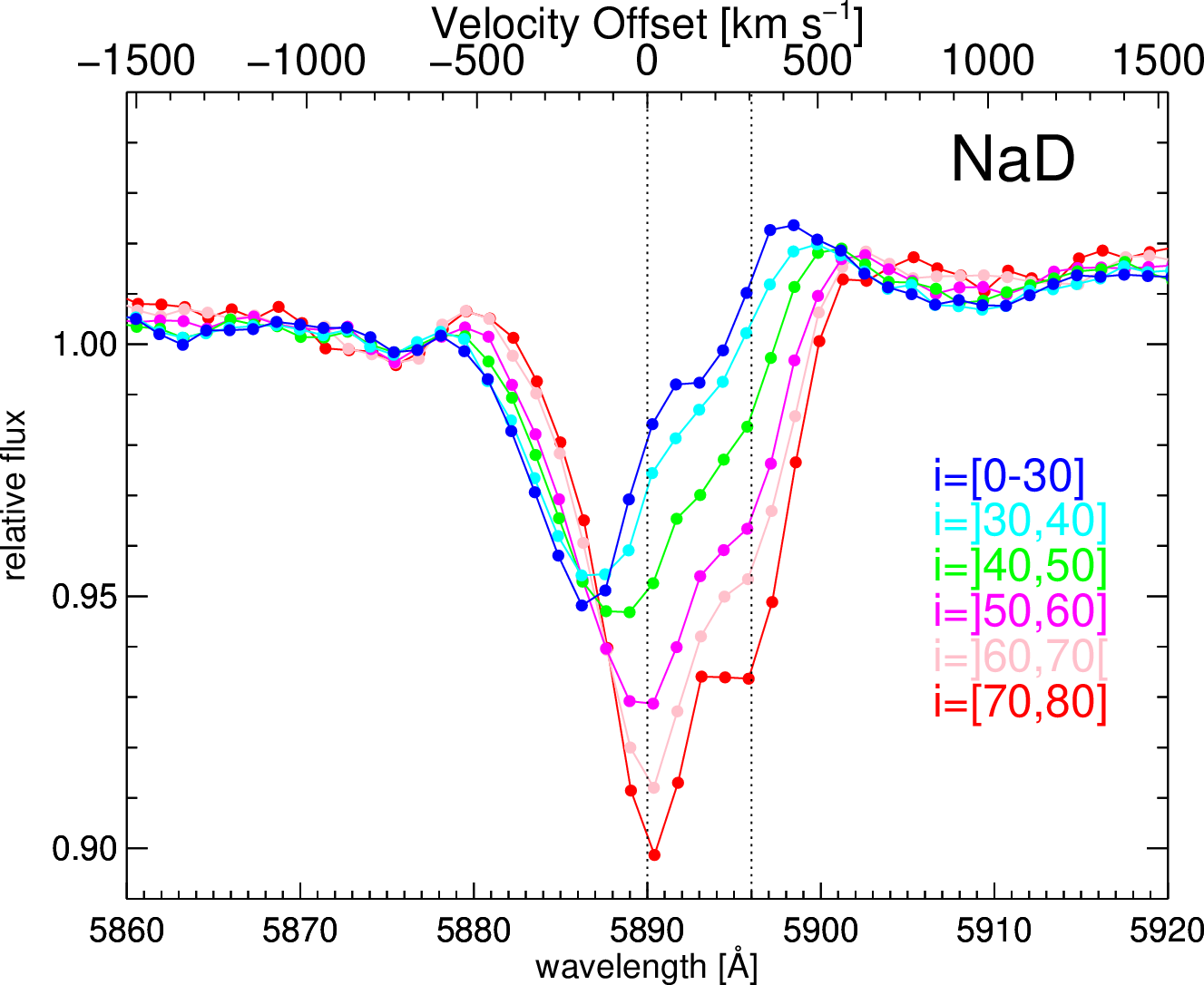}
   \caption{Variation of the ISM Na I D resonant line profile as a function of galaxy inclination, for the galaxies with SFR$\geq 10^{1.1}$ M$_{\sun}/yr$ (including ``pure'' SF galaxies and galaxies with an AGN contribution). The line is centred to the systemic velocity in the edge-on sample and it shows a progressive blue-shift as the mean disk inclination decreases. The maximum shift is reached by face-on galaxies (blue curve).}
              \label{NaD_6iBins}
   \end{figure*}
   \begin{figure}[!t]
   \centering
   \includegraphics[angle=0, width=\hsize ]{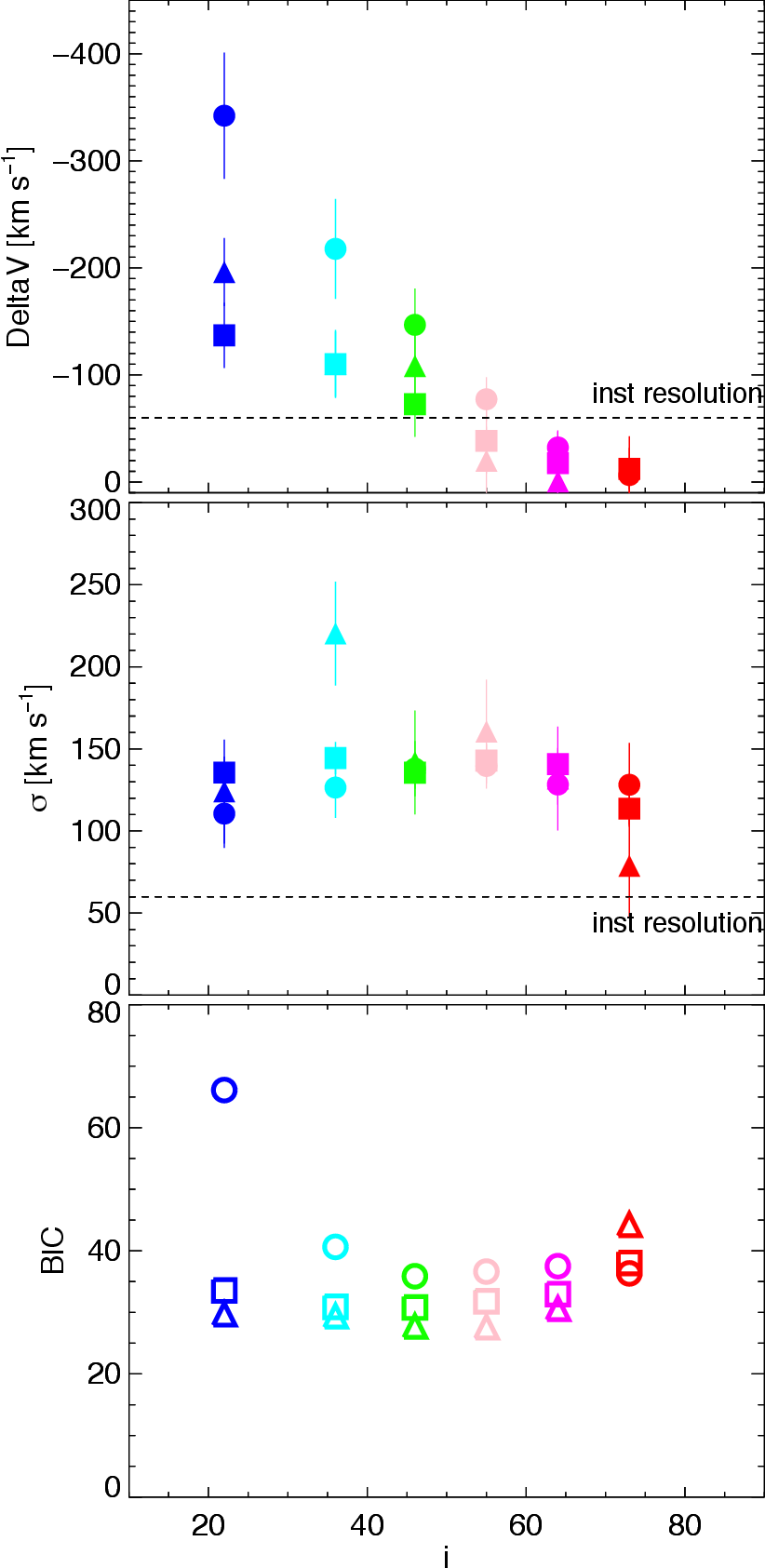}
   \caption{{Na I D velocity shift, velocity dispersion and BIC value estimates with the 3 methods (circles, squares and triangles for method 1, 2 and 3 respectively) described in Section \ref{method}, as a function of the inclination angle, from face-on to edge-on systems (blue and red, respectively). The peak velocity increases towards lower  inclination angles. The width of the outflowing component is reasonably stable for all the 3 methods adopted with a mean value around $\sigma \sim 130 \pm 30$ kms$^{-1}$. The BIC associated with a single doublet component, model 1, is favourite only at large inclination values, while the methods 2 and 3 provide a better fit in the other cases.}}
              \label{NaD_vel_BIC_6iBins}
   \end{figure}

   \begin{figure*}
   \centering
   \includegraphics[angle=90, width=\hsize ]{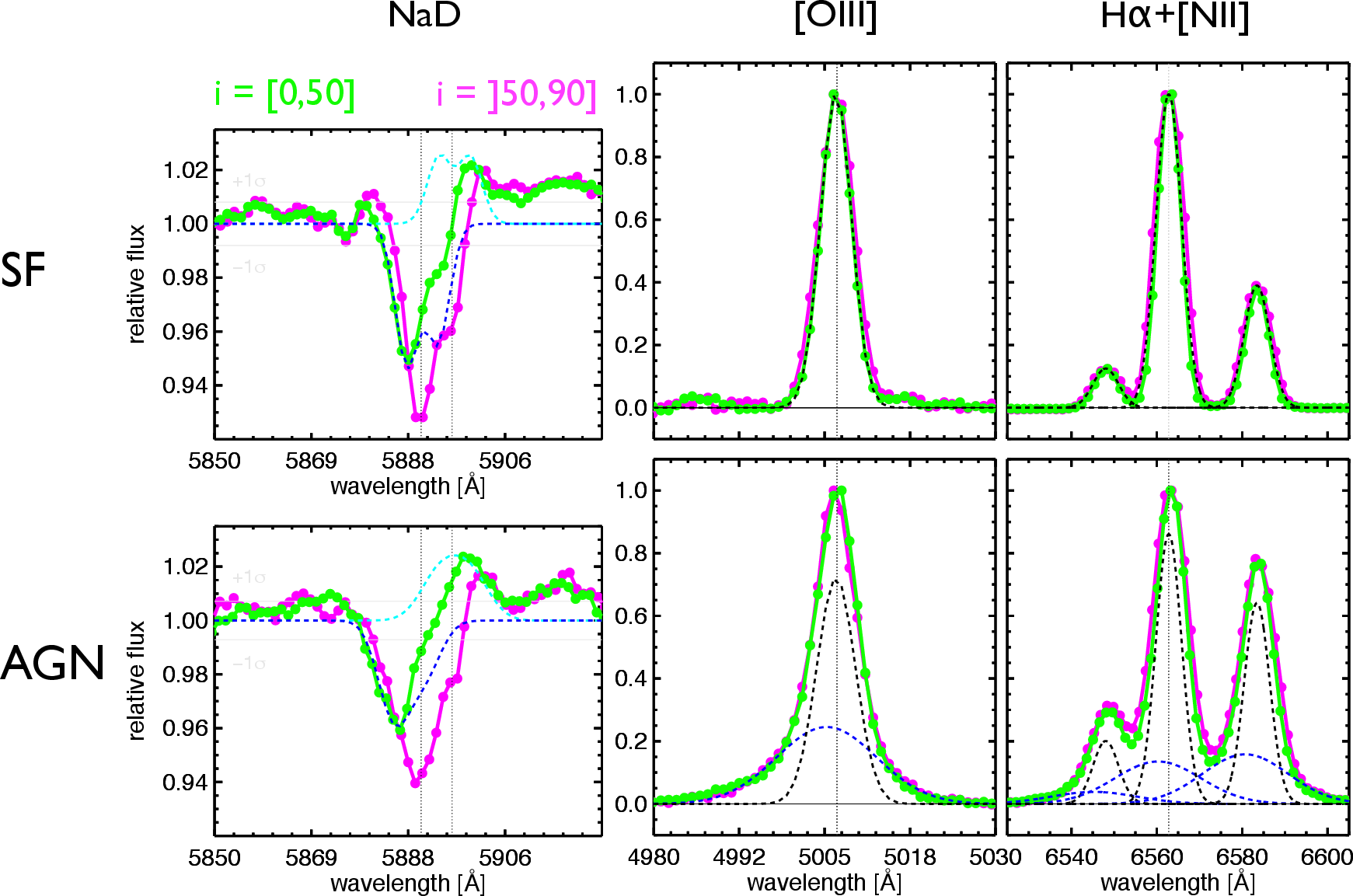}
   \caption{{Correlation between the Na I D resonant line, [OIII]$\lambda 5007$ and H$\alpha$ emission-line profiles (left, central and right panels, respectively) and the galaxy stellar disk inclination (face-on and edge-on systems with inclination $i=[0,50]$ and $i=]50,90]$ in green and magenta, respectively). The spectra are extracted from the ``pure'' SF galaxies (top panels) and for the AGN subsample (bottom panels). The black and blue dashed lines show the decomposition between the narrow systemic and the broad blue-shifted components needed to best fit the data. In the sodium panels, the cyan dashed lines illustrate the emission line component consistently with the P-Cygny predictions showed in the cool gas wind models of \cite{Prochaska+11} and \cite{ScarlataPanagia15}. The Na I D line shift clearly correlates with the galaxy inclination in both SF and AGN samples. The [OIII] and H$\alpha$ emission lines are independent from the galaxy morphology but show a strong shape variation between the SF and AGN subsamples.}}
              \label{NaD_OIII_SF_AGN_inclinazione}
   \end{figure*}

To investigate the dependence of the residual Na I D line profile on the SFR, we stacked all the galaxies with $M{\star}> 5\times 10^{10}$ $M_{\odot}$ in bins of SFR. The results is shown separately for the whole sample, the ``pure'' SF galaxies and the AGN in Fig. \ref{NaD_SFR}. We find that statistically the Na I D residual absorption is observed above the $3\sigma$ level and with a significant blue-shift ($\Delta V > 100$ kms$^{-1}$) only at SFR $\geq 10-12 M_{\sun}/yr$, which is consistent with the starburst region above the local MS (\citealp{RenziniPeng2015}). This is obtained independently of the nuclear activity, because the trend is the same for ``pure'' SF galaxies and AGN hosts separately (central and bottom panels of Fig. \ref{NaD_SFR}, respectively). We point out that, to be conservative, the $\sigma$ level used to estimate the significance of the Na I D residual absorption is estimated in a much larger region of the galaxy spectrum shown in Fig. \ref{NaD_SFR}. This is done to properly take into account the uncertainty of the stellar continuum subtraction.

{{We also point out that the level of significance of the Na I D absorption at level of SFR between 1-10$M_{\odot} yr^{-1}$ could be biased by the contribution of a non negligible percentage of systems with Na I D in emission. Indeed, at $M{\star}$ above $\sim 5\times 10^{10}$ $M_{\odot}$ and $SFR \sim 1-10 M_{\odot} yr^{-1}$ such percentage could be of the order of 30-40\%, as shown in Fig.\ref{fig:FigNEO_NDO}. Thus, emission and absorption contributions could cancel out each other in the stacking, thus, leading to the low significance absorption at SFR between 1-10$M_{\odot} yr^{-1}$ (upper panel of Fig. \ref{NaD_SFR}).}}

\begin{table}
\small
\centering
\begin{tabular}{l|c|c}

\hline
\hline
subsample   & number       & percent              \\
\hline
TOT     &$8772$     & $100\%$        \\
\hline
\hline
i=[0-30]     &$1684$        & $19.2\%$     \\  
i=]30-40]     &$1438$        &$16.4\%$    \\  
i=]40-50]     &$1769$        &$20.2\%$     \\  
i=]50-60]     &$1779$        & $20.3\%$      \\  
i=]60-70[     &$1336$        & $15.2\%$      \\  
i=[70-90]     &$766$        &$8.7\%$   \\  
\hline
\hline
\end{tabular}
\caption{{Basic data about the starburst Sample (galaxies with SFR$> 10^{1.1} M_{\odot}$/yr) slitted in six bins of inclination angle taken from \cite{Simard+11}.}}
\label{tab:InfoSample1}
\end{table}
\normalsize

\subsection{Cold winds in starburst galaxies}\label{starburst}
Based on the results shown in the previous Section, we focus on galaxies located at SFR $\geq 12.5 M_{\sun}/yr$, where there is clear signature of blue-shifted interstellar Na I D absorption. We refer to this sub-sample as the starburst Sample, containing $15,284$ galaxies. {We investigate the relation between significance of the Na I D absorption and its blue-shift with respect to {\it{i)}} galactic disk inclination, to identify the wind geometry, {\it{ii)}} the galaxy nuclear activity to identify the ejection mechanism and, {\it{iii)}} the concurrence with the ionized wind to better study the multiphase nature of the wind. In the following analysis we apply the fitting procedure described in Section \ref{ISMmeasure}.}

\subsubsection{Trends with galaxy inclination}

{In order to study the wind geometry, we split the starburst Sample in six bins of inclination angle taken from the  bulge/disk  decomposition of \cite{Simard+11} catalogue (see Tab. \ref{tab:InfoSample1} for the subsample information).}
Fig. \ref{NaD_6iBins} illustrates the interstellar Na I D absorption profiles for composite spectra in six different inclination bins.
We find a clear trend between the blue-shifted absorption profiles and the galaxy inclination: the neutral sodium line shows
a transition from a strong disk-like component, perfectly centered to the systemic velocity in the edge-on system (inclination $i > 50^\circ$ of the disk rotation axis), to an outflow, blue-shifted component in face-on galaxies ($i < 50^\circ$). This result is perfectly consistent with the finding of \cite{Chen+10} for a subsample of SF galaxies. We point out that in our  edge-on systems the stellar continuum light is strong enough to guarantee a robust background continuum illumination which is essential to measure the ISM absorption.

Fig. \ref{NaD_vel_BIC_6iBins} shows the mean velocity shift (top panel), velocity dispersion (central panel) and relative efficiency of the parameterized models, expressed by the BIC value (bottom panel), as a function of the disk orientation. We find that the Na I D lines in highly inclined galaxies can be reproduced with the $one-free-doublet$ model (filled circles in the figure). The velocity of the line peak is consistent with the systemic velocity of the galaxy ($\Delta V\sim 0$ kms$^{-1}$). In the low inclination spectra, we find a very complicated Na I D line structure. In particular, we find evidence for the presence of a blue-shifted absorption component and a line-emission at roughly the systemic velocity, consistently with the P-Cygni predictions showed in the cool gas wind models for the Mg II and Fe II resonance lines of \cite{Prochaska+11} (Monte Carlo radiative transfer techniques) and  \citealp{ScarlataPanagia15}(semi-analytical line transfer model). For these low inclination galaxies the methods 2 and 3 better reproduce the line shape, as it is shown in the Fig. \ref{NaD_vel_BIC_6iBins} bottom panel by the low BIC values of 2 and 3.
In the face-on galaxy sample the velocity of the line peak reaches a maximum value of $\Delta V\sim 200$ kms$^{-1}$ and maximum velocity of $V_{max} \sim 460$ kms$^{-1}$. The line width do not shows, instead, any dependence on the disk inclination and it has a constant value of $\sigma \sim 130 \pm 30$ kms$^{-1}$(see central panel of Fig. \ref{NaD_vel_BIC_6iBins}). This could be likely due to the stacking procedure, which is averaging out all velocity structure information.

{We find that the trend with the galaxy inclination at fixed SFR is present irrespective of the contribution of the nuclear activity.  In all the ionization classes, SF, SFAGN, AGNSF and AGN, we observe the same trend as shown in Fig. \ref{NaD_6iBins} at SFR$> 10 M_{\odot}yr^{-1}$. The right panels of Fig. \ref{NaD_OIII_SF_AGN_inclinazione} show, in particular, the Na I D line profile for ``pure'' SF galaxies (upper panel) and AGN hosts (lower panel) in two bins of inclination ($i < 50^\circ$, green curve, and $i>50^\circ$, magenta curve, respectively). Highly inclined systems show no velocity shift, while face on SF galaxies and AGN hosts exhibit on average the same velocity blue-shift $\Delta V \sim 200 km/s$.}

\begin{table}
\small
\centering
\begin{tabular}{l|c|c}

\hline
\hline
subsample   & number       & percent              \\
\hline
TOT           &$15,284$     & $100\%$     \\
\hline
\hline
SF             &$3735$       & $24.4\%$      \\
SF-AGN    &$4196$       & $27.4\%$       \\
AGN-SF    &$3912$       & $25.6\%$          \\
AGN          &$872$       & $5.7\%$            \\
unClass    &$2569$       & $16.8\%$        \\
\hline
\hline

\end{tabular}
\caption{{Basic data about the starburst Sample (galaxies with SFR$> 12.5 M_{\odot}yr^{-1}$) slitted in five bins of ionization mechanisms.}}
\label{tab:InfoSample2}
\end{table}
\normalsize

\subsection{Neutral versus Ionized wind}\label{ionezed}  
{To analyze the concurrence of neutral gas and ionized gas winds, we compare here the Na I D absorption line shape with the emission line profile of the [OIII]$\lambda5007$, H$\alpha$, [NII]$\lambda6548$ and [NII]$\lambda6584$, whose perturbed kinematics is generally used as tracer of ionized gas outflows (see e.g. \citealp{Concas+17,Perna+2017b,Mullaney2013,Harrison2014}). Such comparison is done only for the starburst galaxy sample, where we observe clear signature of neutral gas outflow. The analysis is performed as a function of the ionization class and disk inclination.}

{The presence of outflow in the emission line profile is identified through the contribution of a second blue-shifted emission component in addition to a narrow component at the systemic velocity. In this respect, we apply here the same  procedure described in C17. Namely, the [OIII]$\lambda5007$ line profile, in particular, is fitted either with a single Gaussian or with a combination of two Gaussians. The best fit is chosen according to the BIC, as explained in Section \ref{ISMmeasure}. 
The decomposition of the H$\alpha$, [NII]$\lambda6548$ and [NII]$\lambda6584$ emission lines is made more complicated by the proximity of all features. Indeed, the blending of such lines make impossible to carefully constrain the kinematics of the second broad and blue-shifted component.
However, by using the results obtained for the [OIII] line fit (line width and shifts of the different components) and leaving the line amplitude free to vary, we are able to successfully reproduce the H$\alpha$ region (see the line fit in right panel of Fig. \ref{NaD_OIII_SF_AGN_inclinazione}).}
{We point out that the second broad component in the H$\alpha$ emission line could be associated also to photon emission originating from the broad line regions of AGN galaxies. For both the proximity of nitrogen emission lines ad this possible contamination, the study of the  ionized outflows signatures in our SDSS spectra cannot be possible by using the H$\alpha$ emission line region alone.}

{Similarly to what found in C17, we do not observe signature of ionized gas wind in ``pure'' SF galaxies. This is true irrespective of the disk inclination. The upper panels of Fig. \ref{NaD_OIII_SF_AGN_inclinazione} show the stacked [OIII]$\lambda5007$ (central panel) and  H$\alpha$, [NII]$\lambda6548$ and [NII]$\lambda6584$ (right panel) profiles in two bins of inclination ($i < 50^\circ$, green curve, and $i>50^\circ$, magenta curve, respectively). While face-on SF galaxies exhibit clear sign of neutral gas wind, no outflow is ongoing in the ionized gas phase.}

{In addition, we confirm also that the significance of outflow signature of ionized gas increases at increasing AGN contribution, with the highest significance for the AGN class. The overall line width of the [OIII] line increases from a FWHM of $\sim 300$ kms$^{-1}$ (corresponding to $\sigma \sim 127$ kms$^{-1}$ ) for ``pure'' SF galaxies to $\sim 600$ kms$^{-1}$ ($\sigma \sim 257$ kms$^{-1}$) for AGN hosts, at fixed $M{\star}$ and SFR. The same is found for the H$\alpha$, [NII]$\lambda6548$ and [NII]$\lambda6584$ line profiles. The bottom central and right panels of Fig. \ref{NaD_OIII_SF_AGN_inclinazione} clearly indicate the presence of a second broad and blue-shifted Gaussian component in all emission lines of the AGN hosts. However, while the neutral gas wind is observed only in face-on active galaxies, the ionized gas outflow does not show any dependence on the disk inclination in this class of objects.}

{The evidence for multiphase gas wind only in the face-on AGN population is in agreement with the results obtained by \cite{Rupke2005b}, who find agreement between neutral and ionized wind kinematics in Seyferts but not in starbursts systems. Similarly, \cite{Rupke+2017} find high coincidence rate (50\% of the sample) of ionized and neutral gas winds in a sample of 10 nearby Type 1 quasars. Differently, \cite{Villar-Martin2011} do not found any clear evidence for neutral outflows in a subsample of 21 most nearby SDSS QSO2.  We point out that our outcome is not directly in disagreement with the \cite{Villar-Martin2011} findings. The tension between the two results could be doe to the low statistic of the \cite{Villar-Martin2011} sample and/or to the high galactic disk inclination of the 21 QSO.}

\section{Discussion}\label{discussion}
\subsection{Triggering mechanism: AGN or SF activity?}

{As mentioned in the Introduction, in the current paradigm of galaxy formation, two main physical processes are advocated to swipe away the galaxy gas supply: (1) the AGN, and (2) the SF activity itself.
In the ``ejective feedback" scenario, both mechanisms are supposed to be able to eject the gas away through powerful winds, reduce the gas reservoir and therefore stop the growth of the galaxies. From an empirical point of view, the most important issue in our understanding of galactic winds is the relative importance of a SF- and AGN-driven winds (e.g. \citealp{Rupke+Veilleux13}).}

{As reported in previous sections, we find that the neutral gas winds as traced by Na I D absorption lines are connected with the SFR without any correlation with the presence of an optically selected AGN. The neutral gas wind could be originated only in galaxies with very high SF activity. 
In addition, as we showed in Sec.4.2.1, the clear inclination dependence of the neutral winds strongly suggests that such outflowing material is mainly ejected from the galactic disk, where the SF regions are located.}

{Instead, by observing the ionized gas phase in the same galaxy spectra, we find that ionized outflows as traced by [OIII], H$\alpha$, [NII]$\lambda6548$ and [NII]$\lambda6584$ emission lines are present on average only in the AGN sample, while they are absent in the SF sample (central and right panels of Fig.\ref{NaD_OIII_SF_AGN_inclinazione}).  This difference has been first pointed out in \cite{Concas+17} for the [OIII] emission line, and it is strengthened in this work by the analysis of the H$\alpha$ and [NII] emission lines complexes. 
Indeed, the fact that such ionized outflows appears only in AGN systems and they are independent from the galactic disk inclination strongly supports the AGN origin of such winds.}

{We can conclude that in the local Universe, different gas phases of the ISM undergo wind events of different nature: the neutral Na I D outflowing gas is mainly ejected from the galactic disk above a threshold of SF activity, while, the ionized gas outflow traced by the [OIII], H$\alpha$, [NII]$\lambda6548$ and [NII]$\lambda6584$ line asymmetry is likely powered by  AGN feedback.}

{As suggested by several papers (e.g. \citealp{Veilleux2005,RupkeVeilleux2005,RupkeVeilleux2013,Cicone2014,Fluetsch+2018}), the energetics of the outflowing gas are more extreme in galaxies that host a powerful AGN 
compared to the normal starburst systems.
For this reason, a common way to discriminate between the SF- and AGN- driven winds is to compare the kinetic and momentum power ($\dot{E}_{w}$ and $\dot{P}_{w}$, respectively) associated to the winds with the AGN bolometric luminosities and the kinetic power expected to be ascribed to stellar processes (e.g. \citealp{Veilleux2005,Brusa2015}).}

{The kinetic and momentum powers of the galactic winds can be expressed as (see \citealp{CanoDiaz2012, Cresci2015}):}
$$\dot{E}_{w} = \frac{1}{2}\dot{M}_{out}v^{2}_{out} ,$$ 
$$\dot{P}_{w} = \dot{M}_{out}v_{out};$$
{where $v_{out}$ is the velocity of the ejected gas and $\dot{M}_{out}$ is the mass outflow rate.
Following \cite{CanoDiaz2012}, the mass outflow rate could be defined as:}
$$\dot{M}^{[OIII]}_{out}= 0.48 \times \frac{CL_{[OIII]} v_{out}} {n_{e} R_{kpc} 10^{[O/H]}} \; M_{\odot}/yr,$$
{where $C$ is the condensation factor ($C \sim 1$), $L_{[OIII]}$ is the oxygen luminosity associated to the wind component in units of 10$^{41}$ erg s$^{-1}$, 10$^{[O/H]}$ is the gas metallicity in solar units, $n_{e}$ is the electron density and $R_{kpc}$ is the radius of the extended wind in units of kpc.
In the neutral gas counterpart, the mass flow rate can be estimated as:}
$$\dot{M}^{H}_{out}= {N_{H}} \times  {R_{5kpc}} \times  {v_{out\_300}} \;  M_{\odot}/yr ,$$
{where $N_{H}$ is the hydrogen column density in units of 10$^{20}$ cm$^{-2}$, $R_{5kpc}$ is the radius of the extended wind in units of $5$ kpc and, $v_{out\_300}$ is the velocity of the ejected gas in units of $300$ kms$^{-1}$ (see \citealp{Weiner+2009,Perna2015}).}

{In both ionized and neutral gas phase, the estimate of the mass outflow rate depends on several assumptions, in particular: the geometry of the winds (are they biconical, spherical or more complex), their extent (R), and the density of the material embedded on the gas flow (n$_e$ and $N_{H}$).}

{While for electron density recent literature results converge on a typical value of n$_{e} \sim 1000$ cm$^{-3}$ (see \citealp{Perna+2017b,ForsterSchreiber2018}), the hydrogen column density in the neutral gas phase result more difficult to estimate especially in stacked spectra.
In fact, by stacking hundreds different galaxies spectra, we are averaging together galaxies with a wide range of optical depth and covering fraction distributions, that are essential for a correct estimate of $N_{H}$.
In addition, in both the neutral and ionized phase, 
it is impossible to determine the exact extension of our outflowing gas. 
Indeed, our spectra are extracted within a 3" diameter region, so we can provide at most an upper limit to the wind extension and so only a lower or upper limit to the ionized and neutral mass outflow rates, respectively.
Moreover, given the relatively large redshift range of our sample, the physical scale sampled vary of $\sim2$ orders of magnitudes within the single sources, which makes up the stacked spectrum and makes impossible even an order of magnitude estimate of the neutral and ionized powers inside the winds.}
{To measure the dimension and also the geometry of such winds is mandatory to spatially resolve the outflow (see the recent integral field spectroscopy analysis,IFS, of \citealp{Rupke+2017,Venturi+2018}).}

{Nevertheless, the studies based on stacked spectra are essential to infer how common are these outflowing phenomena and how they are related to the galaxies properties. For all these reasons, they represent the crucial step to plan dedicated follow-up for new IFS investigations (e.g. the combined study of ionized wind in local optically selected AGN presented in \citealp{Mullaney2013} and \citealp{Harrison2014}).}

\subsection{Effect on baryon cycle}


{As shown in the previous section, in both neutral and ionized phase, the outflow velocity is relatively small, of the order of $200$ and $500$ kms$^{-1}$for the peak velocity shift and the maximum velocity, respectively. We use the catalog of halo masses of \cite{Yang+07} to retrieve the mean halo mass of the ``pure'' SF and AGN hosts. The region of the SFR-$M{\star}$ diagram considered here, at high stellar masses and SFR, is dominated at 60-70\% by central galaxies of massive halos of $10^{12.5-13}$ M$_{\odot}$, whose escape velocity ($v_{escape} \sim 700-1000 $ km s$^{-1}$) largely exceeds the outflow velocity observed here. Thus, we conclude that, on average, the outflowing gas will never escape the halo potential well. \cite{Cazzoli+16} found very consistent results in the analysis of $\sim$ 50 local LIRGs and ULIRGs.}

{These low outflow velocities suggest that the neutral and ionized phases of the outflowing gas remain bound to the galaxy and likely fall back to the disk. These "light breezes" are consistent with the "gas circulation" scenario proposed by \citeauthor*{FraternaliBinney06} (2006, see also \citealp{Fraternali17} for a review) for the Milky Way. In this scenario, the cold gas driven out of the disk by SF feedback,  mixes with the hot halo gas. The gas mixture cools down through metal lines emission and falls back to the disk with a ballistic trajectory, creating the so called ``galaxy fountain''. In this picture the gas outflow does not stop the galaxy star formation activity but at the contrary it sustains the process by triggering the cooling of the hot coronal gas and so further accretion.}

{We do not find, however, any evidence of accreting recycling gas in any of the analyzed stacked spectra.
This may be related to the predicted low velocities of the inflowing material. For typical spiral galaxies, the fountain gas is expected to fall back at velocities of the order of $70-100$ kms$^{-1}$ or less (as predicted by \citealp{FraternaliBinney06} for NGC 891 and by \citealp{Marasco+12} for the Milky Way), which are comparable to the velocity resolution of the SDSS spectra. Furthermore, as discussed by \cite{Fraternali17}, the fountain gas could fall back at larger radii with respect to the location where the peak of the SFR density occurs. The $3"$ diameter fiber in the SDSS spectra does not allow us to sample the more external region of the galaxies. Finally, recent parsec-scale hydrodynamical simulations, including the presence of thermal conduction, show that the efficiency of the fountain-driven gas accretion strongly decreases with increasing virial temperature of the halo gas \citep{Armillotta+16}. This effect can by particularly important for our massive galaxies which reside in high-mass dark matter halos.}


\section{Summary and Conclusions}\label{conclusions}
{In this paper, we explore the presence of galactic neutral gas winds in a statistically significant sample of local galaxies, drawn from the SDSS, including ``normal'' and ``active'' galaxies, in passive, star-forming and starburst hosts. We use, in particular, the Na I D absorption feature as a tracer cold neutral gas outflow. We interpret the detection of a blueshifted component with respect to the galaxy systemic velocity, as evidence of outflowing gas and we measure its mean and maximum outflow velocity. 
The single SDSS spectra are stacked together in bins of: (a) SFR and M$_{\star}$ to study the neutral gas wind occurrence as a function of the galaxy position with respect to the MS of star forming galaxies, (b) main photoionization mechanism (AGN and SF activity or a combination of the two), to identify the most likely ejecting mechanism (SF or AGN feedback) and, (c) galaxies inclination, to investigate the outflow geometry.
In each stacked spectrum, the stellar continuum is carefully subtracted in order to remove the Na I D stellar absorption component and so enhance the residual ISM component. The main results of the paper may be summarized as follows.}

\begin{itemize}
\item {The neutral gas winds traced by the Na I D line are not ubiquitous in the SFR-M$\star$ plane. A clear signature is detected in massive ($M{\star}> 5 \times 10^{10} M_{\odot}$) highly star forming galaxies (SFR$> 10-12 M_{\odot} yr^{-1}$). However, we do not exclude the presence of outflow signature in less star forming galaxies at fixed mass, where the evidence could be washed out by the stacking of a similar percentage of Na I D in emission and absorption.  In all other loci of the SFR-$M{\star}$ plane the Na I D is observed either in emission (low stellar masses) or in absorption but always at the systemic velocity (high stellar masses and low SFR), without exhibiting any evidence of outflow.}

\item {The same results are obtained irrespective of the BPT classification: the AGN population follows the same trend in the SFR-M$_{\star}$ plane as the SF galaxies population (see Fig.\ref{NaD_SFR}).}

\item {We perform an in depth analysis of the galaxies in the starburst region (SFR $> 10 M_{\odot}$yr$^{-1}$) at very high mass to study the geometry of the outflow. The Na I D residual absorption shows a progressive blue-shift with respect to the systemic velocity with the decrease of the disk inclination. The largest values of peak and maximum velocity shift ($\Delta V \sim 200$ and $V_{max}\sim 460$ kms$^{-1}$, respectively) are observed in the low inclination/ face-on galaxies, while no outflow is observed in edge-on systems. This would suggest that the outflow direction is perpendicular to the disk or it exhibits a large opening angle.}

\item {We find no strong correlation between the evidence of neutral gas wind and the nuclear activity.
The Na I D line profile of ``pure'' SF, SFAGN, AGNSF and AGN samples shows the same velocity shift and the same dependence on disk inclination.}

\item {In the face-on galaxies the Na I D residual component profile is also characterized by a typical P-Cygni profile as predicted by several outflow models (see Prochaska et al. 2011 and Scarlata \& Panagia 2015). This particular feature is also consistent with recent IFU observations of Na I D resonant lines in low redshift QSO systems (e.g. Rupke \& Veilleux 2015 and Rupke et al. 2017). }

\item {``Pure'' SF galaxies do not show any evidence of multiphase gas wind. The observed cold gas outflows traced by the Na I D in galaxies at high stellar mass and SFR is not accompanied by any evidence of ionized gas phase wind, traced by  [OIII]$\lambda5007$, H$\alpha$, [NII]$\lambda6548$ and [NII]$\lambda6584$ emission lines.}

\item {We find clear evidence of multiphase gas outflow only in the AGN population. However, while the Na I D blue-shift strongly depends on the inclination and it is the maximum for face on galaxies, the  [OIII]$\lambda5007$, H$\alpha$, [NII]$\lambda6548$ and [NII]$\lambda6584$ line profiles do not show any dependence of the galactic disk inclination. This suggest that the cold gas outflow is driven by SF feedback above a given SFR threshold, while the ionized gas outflow originates from the nuclear region. Its direction is likely related to the black-hole accretion disk inclination which is decoupled from the host disk.}

\end{itemize}


{We point out that our results do not exclude that strong neutral gas wind might be caused by BH feedback in powerful QSOs, as observed e.g. by \cite{Rupke2005a} and \cite{RupkeVeilleux2011,Rupke+Veilleux13,Rupke&Veilleux2015}. However, such systems are very rare objects in the Local Universe and they do not dominate the mean in the stacking analysis presented in this work. }

{Finally, the low outflow velocities observed in both the neutral and ionized phases suggest that the outflowing gas remain bound to the galaxy and likely fall back to the disk with no definitive effect on gas reservoir and on the galaxy SF activity. 
These ``light breezes" observed in the SDSS galaxy spectra are consistent with the ``gas circulation" scenario (\citealp{FraternaliBinney06} and \citealp{Fraternali17} for a recent review). We do not find, however, any evidence of accreting re-cycling gas expected from a fountain cycle.}

{Much higher spectral ans spatial resolution IFS observations are necessary to better understand the geometry of the moderate outflowing gas observed here, to constrain its energetics, its nature and its effect on the galaxy baryon cycle.}

\begin{acknowledgements}
 We thank the anonymous referee for a constructive and helpful report. AC thanks Federico Lelli for precious suggestions and stimulating discussion. AC, PP and MB are grateful to the Munich Institute for Astrophysics and Particle Physics for its hospitality, when most part of the discussion of this paper was carried out. This research was supported by the DFG cluster of excellence 'Origin and Structure of the Universe'.

\end{acknowledgements}

%
%

%
%

\bibliographystyle{aa}
\bibliography{alice} 

\begin{thebibliography}{117}
\expandafter\ifx\csname natexlab\endcsname\relax\def\natexlab#1{#1}\fi

\bibitem[{{Abazajian} {et~al.}(2009)}]{Abazajian2009}
{Abazajian}, K.~N. {et~al.} 2009, ApJS, 182, 543

\bibitem[{{Alloin} \& {Bica}(1989)}]{Alloin1989}
{Alloin}, D. \& {Bica}, E. 1989, \aap, 217, 57

\bibitem[{{Armillotta} {et~al.}(2016){Armillotta}, {Fraternali}, \&
  {Marinacci}}]{Armillotta+16}
{Armillotta}, L., {Fraternali}, F., \& {Marinacci}, F. 2016, \mnras, 462, 4157

\bibitem[{{Baldry} {et~al.}(2008){Baldry}, {Glazebrook}, \&
  {Driver}}]{Baldry2008}
{Baldry}, I., {Glazebrook}, K., \& {Driver}, S. 2008, MNRAS, 388, 945

\bibitem[{{Baldwin} {et~al.}(1981){Baldwin}, {Phillips}, \& {Terlevich}}]{BPT}
{Baldwin}, J.~A., {Phillips}, M.~M., \& {Terlevich}, R. 1981, \pasp, 93, 5

\bibitem[{{Behroozi} {et~al.}(2010){Behroozi}, {Conroy}, \&
  {Wechsler}}]{Behroozi2010}
{Behroozi}, P.~S., {Conroy}, C., \& {Wechsler}, R.~H. 2010, ApJ, 717, 379

\bibitem[{{Behroozi} {et~al.}(2013){Behroozi}, {Wechsler}, \&
  {Conroy}}]{Behroozi2013}
{Behroozi}, P.~S., {Wechsler}, R.~H., \& {Conroy}, C. 2013, ApJ, 770, 57

\bibitem[{{Bica} \& {Alloin}(1986)}]{Bica1986}
{Bica}, E. \& {Alloin}, D. 1986, \aap, 166, 83

\bibitem[{{Bordoloi} {et~al.}(2014){Bordoloi}, {Lilly}, {Hardmeier}, {Contini},
  {Kneib}, {Le Fevre}, {Mainieri}, {Renzini}, {Scodeggio}, {Zamorani},
  {Bardelli}, {Bolzonella}, {Bongiorno}, {Caputi}, {Carollo}, {Cucciati}, {de
  la Torre}, {de Ravel}, {Garilli}, {Iovino}, {Kampczyk}, {Kova{\v c}},
  {Knobel}, {Lamareille}, {Le Borgne}, {Le Brun}, {Maier}, {Mignoli}, {Oesch},
  {Pello}, {Peng}, {Perez Montero}, {Presotto}, {Silverman}, {Tanaka}, {Tasca},
  {Tresse}, {Vergani}, {Zucca}, {Cappi}, {Cimatti}, {Coppa}, {Franzetti},
  {Koekemoer}, {Moresco}, {Nair}, \& {Pozzetti}}]{Bordoloi+2014}
{Bordoloi}, R., {Lilly}, S.~J., {Hardmeier}, E., {et~al.} 2014, \apj, 794, 130

\bibitem[{{Bower} {et~al.}(2006){Bower}, {Benson}, {Malbon},
  {et~al.}}]{Bower2006}
{Bower}, R.~G., {Benson}, A., {Malbon}, R., {et~al.} 2006, MNRAS, 370, 645

\bibitem[{{Brinchmann} {et~al.}(2004){Brinchmann}, {Charlot}, {White},
  {Tremonti}, {Kauffmann}, {Heckman}, \& {Brinkmann}}]{Brinchmann2004}
{Brinchmann}, J., {Charlot}, S., {White}, S., {et~al.} 2004, MNRAS, 351, 1151

\bibitem[{{Brusa} {et~al.}(2015){Brusa}, {Bongiorno}, {Cresci}, {Perna},
  {et~al.}}]{Brusa2015}
{Brusa}, M., {Bongiorno}, A., {Cresci}, G., {Perna}, M., {et~al.} 2015, MNRAS,
  446, 2394

\bibitem[{{Brusa} {et~al.}(2016){Brusa}, {Perna}, {Cresci}, {Schramm},
  {Delvecchio}, {Lanzuisi}, {Mainieri}, {Mignoli}, {Zamorani}, {Berta},
  {Bongiorno}, {Comastri}, {Fiore}, {Kakkad}, {Marconi}, {Rosario}, {Contini},
  \& {Lamareille}}]{Brusa2016}
{Brusa}, M., {Perna}, M., {Cresci}, G., {et~al.} 2016, \aap, 588, A58

\bibitem[{{Bruzual} \& {Charlot}(2003)}]{BC03}
{Bruzual}, G. \& {Charlot}, S. 2003, MNRAS, 344, 1000

\bibitem[{{Cano-D{\'{\i}}az} {et~al.}(2012){Cano-D{\'{\i}}az}, {Maiolino},
  {Marconi}, {Netzer}, {Shemmer}, \& {Cresci}}]{CanoDiaz2012}
{Cano-D{\'{\i}}az}, M., {Maiolino}, R., {Marconi}, A., {et~al.} 2012, \aap,
  537, L8

\bibitem[{{Cappellari} \& {Emsellem}(2004)}]{Cappelalri_Emsellem2004}
{Cappellari}, M. \& {Emsellem}, E. 2004, PASP, 116, 138

\bibitem[{{Carter} {et~al.}(1986){Carter}, {Visvanathan}, \&
  {Pickles}}]{Carter+1986}
{Carter}, D., {Visvanathan}, N., \& {Pickles}, A.~J. 1986, \apj, 311, 637

\bibitem[{{Cazzoli} {et~al.}(2014){Cazzoli}, {Arribas}, {Colina},
  {Piqueras-L{\'o}pez}, {Bellocchi}, {Emonts}, \& {Maiolino}}]{Cazzoli+14}
{Cazzoli}, S., {Arribas}, S., {Colina}, L., {et~al.} 2014, \aap, 569, A14

\bibitem[{{Cazzoli} {et~al.}(2016){Cazzoli}, {Arribas}, {Maiolino}, \&
  {Colina}}]{Cazzoli+16}
{Cazzoli}, S., {Arribas}, S., {Maiolino}, R., \& {Colina}, L. 2016, \aap, 590,
  A125

\bibitem[{{Chabrier}(2003)}]{Chabrier2003}
{Chabrier}, G. 2003, PASP, 115, 763

\bibitem[{{Chen} {et~al.}(2010){Chen}, {Tremonti}, {Heckman},
  {et~al.}}]{Chen+10}
{Chen}, Y.-M., {Tremonti}, C., {Heckman}, T., {et~al.} 2010, AJ, 140, 445

\bibitem[{{Chevalier}(1977)}]{Chevalier77}
{Chevalier}, R.~A. 1977, \araa, 15, 175

\bibitem[{{Cicone} {et~al.}(2018){Cicone}, {Brusa}, {Ramos Almeida}, {Cresci},
  {Husemann}, \& {Mainieri}}]{Cicone+2018}
{Cicone}, C., {Brusa}, M., {Ramos Almeida}, C., {et~al.} 2018, Nature
  Astronomy, 2, 176

\bibitem[{{Cicone} {et~al.}(2016){Cicone}, {Maiolino}, \&
  {Marconi}}]{Cicone2016}
{Cicone}, C., {Maiolino}, R., \& {Marconi}, A. 2016, A\&A, 588, A41

\bibitem[{{Cicone} {et~al.}(2014){Cicone}, {Maiolino}, {Sturm},
  {et~al.}}]{Cicone2014}
{Cicone}, C., {Maiolino}, R., {Sturm}, E., {et~al.} 2014, A\&A, 562, A21

\bibitem[{{Concas} {et~al.}(2017){Concas}, {Popesso}, {Brusa}, {Mainieri},
  {Erfanianfar}, \& {Morselli}}]{Concas+17}
{Concas}, A., {Popesso}, P., {Brusa}, M., {et~al.} 2017, \aap, 606, A36

\bibitem[{{Conroy} \& {Wechsler}(2009)}]{ConroyWechsler2009}
{Conroy}, C. \& {Wechsler}, R.~H. 2009, ApJ, 696, 620

\bibitem[{{Cresci} \& {Maiolino}(2018)}]{CresciMaiolino2018}
{Cresci}, G. \& {Maiolino}, R. 2018, Nature Astronomy, 2, 179

\bibitem[{{Cresci} {et~al.}(2015)}]{Cresci2015}
{Cresci}, G. {et~al.} 2015, ApJ, 799, 82

\bibitem[{{Croton} {et~al.}(2006)}]{Croton2006}
{Croton}, D., J. {et~al.} 2006, MNRAS, 365, 11

\bibitem[{{Davis} {et~al.}(2012){Davis}, {Krajnovi{\'c}}, {McDermid}, {Bureau},
  {Sarzi}, {Nyland}, {Alatalo}, {Bayet}, {Blitz}, {Bois}, {Bournaud},
  {Cappellari}, {Crocker}, {Davies}, {de Zeeuw}, {Duc}, {Emsellem}, {Khochfar},
  {Kuntschner}, {Lablanche}, {Morganti}, {Naab}, {Oosterloo}, {Scott}, {Serra},
  {Weijmans}, \& {Young}}]{Davis+12}
{Davis}, T.~A., {Krajnovi{\'c}}, D., {McDermid}, R.~M., {et~al.} 2012, \mnras,
  426, 1574

\bibitem[{{De Lucia} {et~al.}(2006){De Lucia}, {Springel}, {White}, {Croton},
  \& {Kauffmann}}]{DeLucia2006}
{De Lucia}, G., {Springel}, V., {White}, S., {Croton}, D., \& {Kauffmann}, G.
  2006, MNRAS, 366, 499

\bibitem[{{Di Matteo} {et~al.}(2005){Di Matteo}, {Springel}, \&
  {Hernquist}}]{DiMatteo2005}
{Di Matteo}, T., {Springel}, V., \& {Hernquist}, L. 2005, nature, 433, 604

\bibitem[{{Erb}(2015)}]{Erb2015}
{Erb}, D. 2015, Nature, 523, 169

\bibitem[{{Fabian}(2012)}]{Fabian2012}
{Fabian}, A. 2012, ARAA, 50, 455

\bibitem[{{Feltre} {et~al.}(2018){Feltre}, {Bacon}, {Tresse}, {Finley},
  {Carton}, {Blaizot}, {Bouch{\'e}}, {Garel}, {Inami}, {Boogaard},
  {Brinchmann}, {Charlot}, {Chevallard}, {Contini}, {Michel-Dansac}, {Mahler},
  {Marino}, {Maseda}, {Richard}, {Schmidt}, \& {Verhamme}}]{Feltre+2018}
{Feltre}, A., {Bacon}, R., {Tresse}, L., {et~al.} 2018, \aap, 617, A62

\bibitem[{{Feruglio} {et~al.}(2015){Feruglio}, {Fiore}, {Carniani},
  {Piconcelli}, {Zappacosta}, {Bongiorno}, {Cicone}, {Maiolino}, {Marconi},
  {Menci}, {Puccetti}, \& {Veilleux}}]{Feruglio+2015}
{Feruglio}, C., {Fiore}, F., {Carniani}, S., {et~al.} 2015, \aap, 583, A99

\bibitem[{{Feruglio} {et~al.}(2010){Feruglio}, {Maiolino}, {Piconcelli},
  {Menci}, {Aussel}, {Lamastra}, \& {Fiore}}]{Feruglio2010}
{Feruglio}, C., {Maiolino}, R., {Piconcelli}, E., {et~al.} 2010, A\&A, 518,
  L155

\bibitem[{{Finley} {et~al.}(2017){Finley}, {Bouch{\'e}}, {Contini}, {Paalvast},
  {Boogaard}, {Maseda}, {Bacon}, {Blaizot}, {Brinchmann}, {Epinat}, {Feltre},
  {Marino}, {Muzahid}, {Richard}, {Schaye}, {Verhamme}, {Weilbacher}, \&
  {Wisotzki}}]{Finley+2017}
{Finley}, H., {Bouch{\'e}}, N., {Contini}, T., {et~al.} 2017, \aap, 608, A7

\bibitem[{{Fluetsch} {et~al.}(2018){Fluetsch}, {Maiolino}, {Carniani},
  {Marconi}, {Cicone}, {Bourne}, {Costa}, {Fabian}, {Ishibashi}, \&
  {Venturi}}]{Fluetsch+2018}
{Fluetsch}, A., {Maiolino}, R., {Carniani}, S., {et~al.} 2018, ArXiv e-prints
  [\eprint[arXiv]{1805.05352}]

\bibitem[{{F{\"o}rster Schreiber} {et~al.}(2018){F{\"o}rster Schreiber},
  {{\"U}bler}, {Davies}, {Genzel}, {Wisnioski}, {Belli}, {Shimizu}, {Lutz},
  {Fossati}, {Herrera-Camus}, {Mendel}, {Tacconi}, {Wilman}, {Beifiori},
  {Brammer}, {Burkert}, {Carollo}, {Davies}, {Eisenhauer}, {Fabricius},
  {Lilly}, {Momcheva}, {Naab}, {Nelson}, {Price}, {Renzini}, {Saglia},
  {Sternberg}, {van Dokkum}, \& {Wuyts}}]{ForsterSchreiber2018}
{F{\"o}rster Schreiber}, N.~M., {{\"U}bler}, H., {Davies}, R.~L., {et~al.}
  2018, ArXiv e-prints [\eprint[arXiv]{1807.04738}]

\bibitem[{{Fraternali}(2017)}]{Fraternali17}
{Fraternali}, F. 2017, in Astrophysics and Space Science Library, Vol. 430,
  Astrophysics and Space Science Library, ed. A.~{Fox} \& R.~{Dav{\'e}}, 323

\bibitem[{{Fraternali} \& {Binney}(2006)}]{FraternaliBinney06}
{Fraternali}, F. \& {Binney}, J.~J. 2006, \mnras, 366, 449

\bibitem[{{Genzel} {et~al.}(2014){Genzel}, {F{\"o}rster Schreiber}, {Rosario},
  {Lang}, {Lutz}, {Wisnioski}, {Wuyts}, {Wuyts}, {Bandara}, {Bender}, {Berta},
  {Kurk}, {Mendel}, {Tacconi}, {Wilman}, {Beifiori}, {Brammer}, {Burkert},
  {Buschkamp}, {Chan}, {Carollo}, {Davies}, {Eisenhauer}, {Fabricius},
  {Fossati}, {Kriek}, {Kulkarni}, {Lilly}, {Mancini}, {Momcheva}, {Naab},
  {Nelson}, {Renzini}, {Saglia}, {Sharples}, {Sternberg}, {Tacchella}, \& {van
  Dokkum}}]{Genzel+14}
{Genzel}, R., {F{\"o}rster Schreiber}, N.~M., {Rosario}, D., {et~al.} 2014,
  \apj, 796, 7

\bibitem[{{Greene} \& {Ho}(2005)}]{GreeneHo2005a}
{Greene}, J. \& {Ho}, L. 2005, ApJ, 627, 721

\bibitem[{{Guo} {et~al.}(2010){Guo}, {White}, {Li}, \&
  {Boylan-Kolchin}}]{Guo2010}
{Guo}, Q., {White}, S., {Li}, C., \& {Boylan-Kolchin}, M. 2010, MNRAS, 404,
  1111

\bibitem[{{Harrison} {et~al.}(2014){Harrison}, {Alexander}, {Mullaney}, \&
  {Swinbank}}]{Harrison2014}
{Harrison}, C., {Alexander}, D.~M., {Mullaney}, J.~R., \& {Swinbank}, A.~M.
  2014, MNRAS, 441, 3306

\bibitem[{{Harrison} {et~al.}(2016){Harrison}, {Alexander}, {Mullaney},
  {Stott}, {Swinbank}, {Arumugam}, {Bauer}, {Bower}, {Bunker}, \&
  {Sharples}}]{Harrison+16}
{Harrison}, C.~M., {Alexander}, D.~M., {Mullaney}, J.~R., {et~al.} 2016,
  \mnras, 456, 1195

\bibitem[{{Heckman} {et~al.}(1990){Heckman}, {Armus}, \& {Miley}}]{Heckman+90}
{Heckman}, T., {Armus}, L., \& {Miley}, G. 1990, ApJS, 74, 833

\bibitem[{{Heckman} {et~al.}(2000){Heckman}, {Lehnert}, {Strickland}, \&
  {Armus}}]{Heckman2000}
{Heckman}, T.~M., {Lehnert}, M.~D., {Strickland}, D.~K., \& {Armus}, L. 2000,
  \apjs, 129, 493

\bibitem[{{Henriques} {et~al.}(2017){Henriques}, {White}, {Thomas}, {Angulo},
  {Guo}, {Lemson}, \& {Wang}}]{Henriques+17}
{Henriques}, B.~M.~B., {White}, S.~D.~M., {Thomas}, P.~A., {et~al.} 2017,
  \mnras, 469, 2626

\bibitem[{{Hill} \& {Zakamska}(2014)}]{HillZakamska2014}
{Hill}, M.~J. \& {Zakamska}, N. 2014, MNRAS, 439, 2701

\bibitem[{{Hopkins} {et~al.}(2006)}]{Hopkins2006}
{Hopkins}, P. {et~al.} 2006, ApJS, 163, 50

\bibitem[{{Hopkins} {et~al.}(2014){Hopkins}, {Kere{\v s}}, {O{\~n}orbe},
  {Faucher-Gigu{\`e}re}, {Quataert}, {Murray}, \& {Bullock}}]{Hopkins+14}
{Hopkins}, P.~F., {Kere{\v s}}, D., {O{\~n}orbe}, J., {et~al.} 2014, \mnras,
  445, 581

\bibitem[{{Jeong} {et~al.}(2013){Jeong}, {Yi}, {Kyeong}, {Sarzi}, {Sung}, \&
  {Oh}}]{Jeong+13}
{Jeong}, H., {Yi}, S.~K., {Kyeong}, J., {et~al.} 2013, \apjs, 208, 7

\bibitem[{{Kauffmann} {et~al.}(2003){Kauffmann}, {Heckman}, {White}, {Charlot},
  {Tremonti}, {Peng}, {Seibert}, {Brinkmann}, {Nichol}, {SubbaRao}, \&
  {York}}]{Kauffmann2003a}
{Kauffmann}, G., {Heckman}, T.~M., {White}, S.~D.~M., {et~al.} 2003, \mnras,
  341, 54

\bibitem[{{Kewley} {et~al.}(2006){Kewley}, {Groves}, {Kauffmann}, \&
  {Heckman}}]{Kewley2006}
{Kewley}, L., {Groves}, B., {Kauffmann}, G., \& {Heckman}, T. 2006, MNRAS, 372,
  961

\bibitem[{{King} \& {Pounds}(2015)}]{KingPounds2015}
{King}, A. \& {Pounds}, K. 2015, \araa, 53, 115

\bibitem[{{Liddle}(2007)}]{Liddle2007}
{Liddle}, A.~R. 2007, \mnras, 377, L74

\bibitem[{{Madau} {et~al.}(1996){Madau}, {Ferguson}, {Dickinson}, {Giavalisco},
  {Steidel}, \& {Fruchter}}]{Madau1996}
{Madau}, P., {Ferguson}, H.~C., {Dickinson}, M., {et~al.} 1996, MNRAS, 283,
  1388

\bibitem[{{Maiolino} {et~al.}(2012){Maiolino}, {Gallerani}, {Neri},
  {et~al.}}]{Maiolino2012}
{Maiolino}, R., {Gallerani}, S., {Neri}, R., {et~al.} 2012, MNRAS, 425, L66

\bibitem[{{Maiolino} {et~al.}(2017){Maiolino}, {Russell}, {Fabian}, {Carniani},
  {Gallagher}, {Cazzoli}, {Arribas}, {Belfiore}, {Bellocchi}, {Colina},
  {Cresci}, {Ishibashi}, {Marconi}, {Mannucci}, {Oliva}, \&
  {Sturm}}]{Maiolino+2017}
{Maiolino}, R., {Russell}, H.~R., {Fabian}, A.~C., {et~al.} 2017, \nat, 544,
  202

\bibitem[{{Marasco} {et~al.}(2012){Marasco}, {Fraternali}, \&
  {Binney}}]{Marasco+12}
{Marasco}, A., {Fraternali}, F., \& {Binney}, J.~J. 2012, \mnras, 419, 1107

\bibitem[{{Martin}(2005)}]{Martin2005}
{Martin}, C. 2005, ApJ, 621, 227

\bibitem[{{Martin}(2006)}]{Martin2006}
{Martin}, C. 2006, ApJ, 647, 222

\bibitem[{{Martin} {et~al.}(2012{\natexlab{a}})}]{Martin+12}
{Martin}, C. {et~al.} 2012{\natexlab{a}}, ApJ, 760, 127

\bibitem[{{Martin} {et~al.}(2012{\natexlab{b}}){Martin}, {Shapley}, {Coil},
  {Kornei}, {Bundy}, {Weiner}, {Noeske}, \& {Schiminovich}}]{Martin2012}
{Martin}, C.~L., {Shapley}, A.~E., {Coil}, A.~L., {et~al.} 2012{\natexlab{b}},
  \apj, 760, 127

\bibitem[{{Moster} {et~al.}(2013){Moster}, {naab}, \& {White}}]{Moster2013}
{Moster}, B.~P., {naab}, T., \& {White}, S. 2013, MNRAS, 428, 3121

\bibitem[{{Moster} {et~al.}(2010){Moster}, {Somerville}, {Maulbetsch},
  {et~al.}}]{Moster2010}
{Moster}, B.~P., {Somerville}, R., {Maulbetsch}, C., {et~al.} 2010, ApJ, 710,
  903

\bibitem[{{Mullaney} {et~al.}(2013){Mullaney}, {Alexander}, {Fine},
  {et~al.}}]{Mullaney2013}
{Mullaney}, J.~R., {Alexander}, D.~M., {Fine}, S., {et~al.} 2013, MNRAS, 433,
  622

\bibitem[{{Murray} {et~al.}(2005){Murray}, {Quataert}, \&
  {Thompson}}]{Murray+05}
{Murray}, N., {Quataert}, E., \& {Thompson}, T.~A. 2005, \apj, 618, 569

\bibitem[{{O'Connell}(1976)}]{O'Connell1976}
{O'Connell}, R.~W. 1976, \apj, 206, 370

\bibitem[{{Parikh} {et~al.}(2018){Parikh}, {Thomas}, {Maraston}, {Westfall},
  {Goddard}, {Lian}, {Meneses-Goytia}, {Jones}, {Vaughan}, {Andrews},
  {Bershady}, {Bizyaev}, {Brinkmann}, {Brownstein}, {Bundy}, {Drory},
  {Emsellem}, {Law}, {Newman}, {Roman-Lopes}, {Wake}, {Yan}, \&
  {Zheng}}]{Parikh+2018}
{Parikh}, T., {Thomas}, D., {Maraston}, C., {et~al.} 2018, \mnras, 477, 3954

\bibitem[{{Perna} {et~al.}(2015){Perna}, {Brusa}, {Salvato}, {Cresci},
  {Lanzuisi}, {Berta}, {Delvecchio}, {et~al.}}]{Perna2015}
{Perna}, M., {Brusa}, M., {Salvato}, M., {et~al.} 2015, A\&A, 583, A72

\bibitem[{{Perna} {et~al.}(2017{\natexlab{a}}){Perna}, {Lanzuisi}, {Brusa},
  {Cresci}, \& {Mignoli}}]{Perna+2017b}
{Perna}, M., {Lanzuisi}, G., {Brusa}, M., {Cresci}, G., \& {Mignoli}, M.
  2017{\natexlab{a}}, ArXiv e-prints [\eprint[arXiv]{1705.08388}]

\bibitem[{{Perna} {et~al.}(2017{\natexlab{b}}){Perna}, {Lanzuisi}, {Brusa},
  {Mignoli}, \& {Cresci}}]{Perna+2017a}
{Perna}, M., {Lanzuisi}, G., {Brusa}, M., {Mignoli}, M., \& {Cresci}, G.
  2017{\natexlab{b}}, \aap, 603, A99

\bibitem[{{Peterson}(1976)}]{Peterson1976}
{Peterson}, R.~C. 1976, \apjl, 210, L123

\bibitem[{{Pettini} {et~al.}(2000){Pettini}, {Steidel}, {Adelberger},
  {Dickinson}, \& {Giovalisco}}]{Pettini2000}
{Pettini}, M., {Steidel}, C., {Adelberger}, K., {Dickinson}, M., \&
  {Giovalisco}, M. 2000, ApJ, 528, 96

\bibitem[{{Prochaska} {et~al.}(2011){Prochaska}, {Kasen}, \&
  {Rubin}}]{Prochaska+11}
{Prochaska}, J.~X., {Kasen}, D., \& {Rubin}, K. 2011, \apj, 734, 24

\bibitem[{{Renzini} \& {Peng}(2015)}]{RenziniPeng2015}
{Renzini}, A. \& {Peng}, Y.-j. 2015, \apjl, 801, L29

\bibitem[{{Rodr{\'{\i}}guez Zaur{\'{\i}}n} {et~al.}(2013){Rodr{\'{\i}}guez
  Zaur{\'{\i}}n}, {Tadhunter}, {Rose}, \& {Holt}}]{Rodriguez-Zaurin+13}
{Rodr{\'{\i}}guez Zaur{\'{\i}}n}, J., {Tadhunter}, C.~N., {Rose}, M., \&
  {Holt}, J. 2013, \mnras, 432, 138

\bibitem[{{Rubin} {et~al.}(2014){Rubin}, {Prochaska}, {Koo},
  {et~al.}}]{Rubin+14}
{Rubin}, K. H.~R., {Prochaska}, J.~X., {Koo}, D.~C., {et~al.} 2014, ApJ, 794,
  156

\bibitem[{{Rupke} {et~al.}(2017){Rupke}, {G{\"u}ltekin}, \&
  {Veilleux}}]{Rupke+2017}
{Rupke}, D., {G{\"u}ltekin}, K., \& {Veilleux}, S. 2017, ArXiv e-prints
  [\eprint[arXiv]{1708.05139}]

\bibitem[{{Rupke} \& {Veilleux}(2011)}]{RupkeVeilleux2011}
{Rupke}, D. \& {Veilleux}, S. 2011, ApJL, 729, L27

\bibitem[{{Rupke} \& {Veilleux}(2013{\natexlab{a}})}]{Rupke+Veilleux13}
{Rupke}, D. \& {Veilleux}, S. 2013{\natexlab{a}}, ApJ, 768, 75

\bibitem[{{Rupke} \& {Veilleux}(2013{\natexlab{b}})}]{RupkeVeilleux2013}
{Rupke}, D. \& {Veilleux}, S. 2013{\natexlab{b}}, ApJL, 775, L15

\bibitem[{{Rupke} {et~al.}(2002){Rupke}, {Veilleux}, \& {Sanders}}]{Rupke2002}
{Rupke}, D., {Veilleux}, S., \& {Sanders}, D. 2002, ApJ, 570, 588

\bibitem[{{Rupke} {et~al.}(2005a){Rupke}, {Veilleux}, \&
  {Sanders}}]{Rupke2005a}
{Rupke}, D., {Veilleux}, S., \& {Sanders}, D. 2005a, ApJ, 160, 87

\bibitem[{{Rupke} {et~al.}(2005b){Rupke}, {Veilleux}, \&
  {Sanders}}]{Rupke2005b}
{Rupke}, D., {Veilleux}, S., \& {Sanders}, D. 2005b, ApJ, 160, 115

\bibitem[{{Rupke} {et~al.}(2005c){Rupke}, {Veilleux}, \&
  {Sanders}}]{Rupke2005c}
{Rupke}, D., {Veilleux}, S., \& {Sanders}, D. 2005c, ApJ, 632, 751

\bibitem[{{Rupke} \& {Veilleux}(2005)}]{RupkeVeilleux2005}
{Rupke}, D.~S. \& {Veilleux}, S. 2005, \apjl, 631, L37

\bibitem[{{Rupke} \& {Veilleux}(2015)}]{Rupke&Veilleux2015}
{Rupke}, D.~S.~N. \& {Veilleux}, S. 2015, \apj, 801, 126

\bibitem[{{Salim} {et~al.}(2007)}]{Salim2007}
{Salim}, J. {et~al.} 2007, ApJS, 173, 267

\bibitem[{{Sarzi} {et~al.}(2006)}]{Sarzi2006}
{Sarzi}, M. {et~al.} 2006, MNRAS, 366, 1151

\bibitem[{{Scarlata} \& {Panagia}(2015)}]{ScarlataPanagia15}
{Scarlata}, C. \& {Panagia}, N. 2015, \apj, 801, 43

\bibitem[{{Schawinski} {et~al.}(2010){Schawinski}, {Urry}, {Virani}, {Coppi},
  {Bamford}, {Treister}, {Lintott}, {Sarzi}, {Keel}, {Kaviraj}, {Cardamone},
  {Masters}, {Ross}, {Andreescu}, {Murray}, {Nichol}, {Raddick}, {Slosar},
  {Szalay}, {Thomas}, \& {Vandenberg}}]{Schawinski+2010}
{Schawinski}, K., {Urry}, C.~M., {Virani}, S., {et~al.} 2010, \apj, 711, 284

\bibitem[{{Simard} {et~al.}(2011){Simard}, {Mendel}, {Patton}, {Ellison}, \&
  {McConnachie}}]{Simard+11}
{Simard}, L., {Mendel}, J.~T., {Patton}, D.~R., {Ellison}, S.~L., \&
  {McConnachie}, A.~W. 2011, \apjs, 196, 11

\bibitem[{{Spiniello} {et~al.}(2012){Spiniello}, {Trager}, {Koopmans}, \&
  {Chen}}]{Spiniello+12}
{Spiniello}, C., {Trager}, S.~C., {Koopmans}, L.~V.~E., \& {Chen}, Y.~P. 2012,
  \apjl, 753, L32

\bibitem[{{Spitzer}(1978)}]{Spitzer1978}
{Spitzer}, Jr., L. 1978, \jrasc, 72, 349

\bibitem[{{Stasinska} {et~al.}(2006){Stasinska}, {Cid Fernandes},
  {et~al.}}]{Stasinska2006}
{Stasinska}, G., {Cid Fernandes}, R., {et~al.} 2006, MNRAS, 371, 972

\bibitem[{{Steidel} {et~al.}(2010){Steidel}, {Erb}, {Shapley}, {Pettini},
  {Reddy}, {Bogosavljevi{\'c}}, {Rudie}, \& {Rakic}}]{Steidel+10}
{Steidel}, C.~C., {Erb}, D.~K., {Shapley}, A.~E., {et~al.} 2010, \apj, 717, 289

\bibitem[{{Strauss} {et~al.}(2002)}]{Strauss2002}
{Strauss}, M.~A. {et~al.} 2002, AJ, 124, 1810

\bibitem[{{Thomas} {et~al.}(2003){Thomas}, {Maraston}, \&
  {Bender}}]{Thomas+2003}
{Thomas}, D., {Maraston}, C., \& {Bender}, R. 2003, \mnras, 343, 279

\bibitem[{{Tremonti} {et~al.}(2004){Tremonti}, {Heckman}, {Kauffmann},
  {Brinchmann}, {Charlot}, {White}, {Seibert}, {Peng}, {Schlegel}, {Uomoto},
  {Fukugita}, \& {Brinkmann}}]{Tremonti+04}
{Tremonti}, C.~A., {Heckman}, T.~M., {Kauffmann}, G., {et~al.} 2004, \apj, 613,
  898

\bibitem[{{van Dokkum} \& {Conroy}(2010)}]{vanDokkum+10}
{van Dokkum}, P.~G. \& {Conroy}, C. 2010, \nat, 468, 940

\bibitem[{{van Dokkum} \& {Conroy}(2012)}]{vanDokkum+12}
{van Dokkum}, P.~G. \& {Conroy}, C. 2012, \apj, 760, 70

\bibitem[{{Veilleux} {et~al.}(2005){Veilleux}, {Cecil}, \&
  {Bland-Hawthorn}}]{Veilleux2005}
{Veilleux}, S., {Cecil}, G., \& {Bland-Hawthorn}, J. 2005, ARAA, 43, 769

\bibitem[{{Venturi} {et~al.}(2018){Venturi}, {Nardini}, {Marconi}, {Carniani},
  {Mingozzi}, {Cresci}, {Mannucci}, {Risaliti}, {Maiolino}, {Balmaverde},
  {Bongiorno}, {Brusa}, {Capetti}, {Cicone}, {Ciroi}, {Feruglio}, {Fiore},
  {Gallazzi}, {La Franca}, {Mainieri}, {Matsuoka}, {Nagao}, {Perna},
  {Piconcelli}, {Sani}, {Tozzi}, \& {Zibetti}}]{Venturi+2018}
{Venturi}, G., {Nardini}, E., {Marconi}, A., {et~al.} 2018, ArXiv e-prints
  [\eprint[arXiv]{1809.01206}]

\bibitem[{{Villar Mart{\'{\i}}n} {et~al.}(2014){Villar Mart{\'{\i}}n},
  {Emonts}, {Humphrey}, {Cabrera Lavers}, \& {Binette}}]{Villar-Martin2014}
{Villar Mart{\'{\i}}n}, M., {Emonts}, B., {Humphrey}, A., {Cabrera Lavers}, A.,
  \& {Binette}, L. 2014, \mnras, 440, 3202

\bibitem[{{Villar-Mart{\'{\i}}n} {et~al.}(2011){Villar-Mart{\'{\i}}n},
  {Humphrey}, {Delgado}, {Colina}, \& {Arribas}}]{Villar-Martin2011}
{Villar-Mart{\'{\i}}n}, M., {Humphrey}, A., {Delgado}, R.~G., {Colina}, L., \&
  {Arribas}, S. 2011, \mnras, 418, 2032

\bibitem[{{Weiner} {et~al.}(2009){Weiner}, {Coil}, {Prochaska}, {Newman},
  {Cooper}, {Bundy}, {Conselice}, {Dutton}, {Faber}, {Koo}, {Lotz}, {Rieke}, \&
  {Rubin}}]{Weiner+2009}
{Weiner}, B.~J., {Coil}, A.~L., {Prochaska}, J.~X., {et~al.} 2009, \apj, 692,
  187

\bibitem[{{Worthey}(1998)}]{Worthey1998}
{Worthey}, G. 1998, \pasp, 110, 888

\bibitem[{{Worthey} {et~al.}(1994){Worthey}, {Faber}, {Gonzalez}, \&
  {Burstein}}]{Worthey+1994}
{Worthey}, G., {Faber}, S.~M., {Gonzalez}, J.~J., \& {Burstein}, D. 1994,
  \apjs, 94, 687

\bibitem[{{Worthey} {et~al.}(2011){Worthey}, {Ingermann}, \&
  {Serven}}]{Worthey2011}
{Worthey}, G., {Ingermann}, B.~A., \& {Serven}, J. 2011, \apj, 729, 148

\bibitem[{{Yang} {et~al.}(2007){Yang}, {Mo}, {van den Bosch}, {Pasquali}, {Li},
  \& {Barden}}]{Yang+07}
{Yang}, X., {Mo}, H.~J., {van den Bosch}, F.~C., {et~al.} 2007, \apj, 671, 153

\bibitem[{{York} {et~al.}(2000)}]{York2000}
{York}, D.~G. {et~al.} 2000, Aj, 120, 1579

\bibitem[{{Zakamska} {et~al.}(2016){Zakamska}, {Hamann}, {P{\^a}ris}, {Brandt},
  {Greene}, {Strauss}, {Villforth}, {Wylezalek}, {Alexandroff}, \&
  {Ross}}]{Zakamska+2016}
{Zakamska}, N.~L., {Hamann}, F., {P{\^a}ris}, I., {et~al.} 2016, \mnras, 459,
  3144

\end{thebibliography}

\end{document}